\documentclass[journal]{IEEEtran}    
\usepackage{amsmath,amsfonts}
\usepackage{algorithmic}
\usepackage{array}
\usepackage{textcomp}
\usepackage{stfloats}
\usepackage{url}
\usepackage{verbatim}
\usepackage{graphicx}
\def\BibTeX{{\rm B\kern-.05em{\sc i\kern-.025em b}\kern-.08em
    T\kern-.1667em\lower.7ex\hbox{E}\kern-.125emX}}
\usepackage{balance}

\usepackage[utf8]{inputenc}
\usepackage[english]{babel}
\usepackage{graphicx}
\usepackage{float}
\usepackage{amssymb}
\usepackage{amsmath}
\usepackage [autostyle, english = american]{csquotes}
\usepackage{multirow}
\usepackage{times}
\usepackage{colortbl}
\usepackage{amsthm}
\usepackage{amsfonts}
\usepackage{array,booktabs,tabularx}
\newcolumntype{M}[1]{>{\centering\arraybackslash}m{#1}}
\DeclareMathAlphabet{\pazocal}{OMS}{zplm}{m}{n}
\MakeOuterQuote{"}
\usepackage[bookmarks=false]{hyperref}

\usepackage{capt-of} 

\usepackage{dsfont}

\usepackage{graphicx}
\usepackage{mathtools}
\usepackage[ruled,  linesnumbered]{algorithm2e}
\usepackage{subcaption}

\usepackage{xcolor}
\usepackage{etoolbox}
\usepackage[export]{adjustbox}

\usepackage{lipsum}
\usepackage{mathtools}
\usepackage{cuted}

\usepackage{epsf}

\begin{document}
\title{Multi-Agent Reinforcement Learning for Distributed Resource Allocation in Cell-Free Massive MIMO-enabled Mobile Edge Computing Network}

\author{Fitsum Debebe Tilahun, 
	Ameha Tsegaye Abebe, 
	and~Chung G. Kang,~\IEEEmembership{Senior~Member,~IEEE}}
	
\author{Fitsum Debebe Tilahun, 
	Ameha Tsegaye Abebe, 
	and~Chung G. Kang,~\IEEEmembership{Senior~Member,~IEEE}
\thanks{
This work was supported by the National Research Foundation of Korea (NRF) grant funded by the Korean government (MSIT) (No.2020R1A2C100998413). ({\it Corresponding author: Chung G. Kang})
Fitsum Debebe Tilahun and Chung G. Kang are with the School
of Electrical Engineering, Korea University, Seoul,
South Korea (e-mail: fitsum\_debebe@korea.ac.kr, ccgkang@korea.ac.kr). 
Ameha Tsegaye Abebe is with Samsung Research, Seoul, South Korea (e-mail: amehat.abebe@samsung.com).  A part of this work was presented at the IEEE Global Communications Conference (GLOBECOM) 2022 [1].  
}}


\maketitle

\begin{abstract}
To support the newly introduced multimedia services with ultra-low latency and extensive computation requirements, resource-constrained end-user devices should utilize the ubiquitous computing resources available at network edge for augmenting on-board (local) processing with edge computing. In this regard, the capability of cell-free massive MIMO to provide reliable access links by guaranteeing uniform quality of service without cell edge can be exploited for a seamless parallel computing. Taking this into account, we formulate a joint communication and computing resource allocation (JCCRA) problem for a cell-free massive MIMO-enabled mobile edge computing (MEC) network with the objective of minimizing the total energy consumption of the users while meeting the ultra-low delay constraints. To derive efficient and adaptive JCCRA scheme robust to network dynamics, we present a distributed solution approach based on cooperative multi-agent reinforcement learning. The simulation results demonstrate that the proposed distributed approach can achieve comparable performance to a centralized deep deterministic policy gradient (DDPG)-based target benchmark, without incurring additional overhead and time cost. It is also shown that our approach significantly outperforms heuristic baselines in terms of energy efficiency, roughly up to 5 times less total energy consumption. Furthermore, we demonstrate substantial performance improvement compared to cellular MEC systems. 
\end{abstract}

\begin{IEEEkeywords}
Joint communication and computing resource allocation (JCCRA), mobile edge computing, cell-free massive MIMO, multi-agent reinforcement learning.
\end{IEEEkeywords}

\section{Introduction}
{\IEEEPARstart {T}{he} \textcolor{black}{past few years have seen a rapid increase in computationally intensive applications such as face recognition, real time gaming, autonomous driving and so on.} One key aspect of the services is their tight delay requirement which poses a serious challenge to user equipment (UE) with limited battery power, computational and storage capabilities. To handle these bottlenecks at the UEs, computation offloading to powerful cloud computing platforms had been implemented for various services \textcolor{black}{[2], [3]}. Recently, a paradigm that has brought computing resources to the edge of the network, dubbed as mobile edge computing (MEC), has been introduced to further reduce the experienced latency. As we move towards the next decade, however, the current cellular MEC systems can hardly keep up with the diverse and yet, even more stringent requirements of the envisaged next generation advanced services, and the evolution of the existing ones \textcolor{black}{[4], [5]}. For instance, expected use cases of the sixth generation (6G) communication network \textcolor{black}{[6]}, such as multi-sensory extended reality (augmented, virtual, and mixed reality), and holographic displays, require extremely high data rate, improved coverage and spectral efficiency, deterministic ultra-low latency, and extensive computational capability \textcolor{black}{[7]}. 

To attain the requirements of the envisioned applications, however, a functionality of reliable \emph{virtual bus} might be required for parallel \& cooperative processing, i.e., to connect the users with ubiquitous computing resources at the edge, such as nearby mobile users (with idle or more powerful processors), and edge/cloud servers (with full-blown computing resources). In this regard, we consider a cell-free massive MIMO system \textcolor{black}{[8]}, one of the potential network infrastructures envisioned for beyond-5G and 6G networks, for enabling a seamless computation offloading. A cell-free network can provide sufficiently fast and reliable access throughout the coverage of the network, virtually eliminating cell-edge users. This is accomplished by serving a relatively small number of users simultaneously from several geographically distributed access points (APs) that are connected to a central processing unit (CPU) \textcolor{black}{[9]}. In the mobile edge computing network under consideration, the CPU is equipped with an edge server of finite capacity to render computing services for resource-constrained users with time-critical and computationally intensive tasks. As opposed to the unreliable wireless links in cellular MEC systems, the cell-free access links in our framework can play the key functionality of realizing virtual buses for parallel computing, i.e., allowing for instant access to the edge server in the CPU, irrespective of users’ location, thus \textcolor{black}{providing energy-efficient and consistently low-latency computational task offloading.}

In this paper, we formulate a joint communication and computing resource allocation (JCCRA) problem for the cell-free massive MIMO-enabled mobile edge computing network. Specifically, with the objective of minimizing the total energy consumption of the users subject to the ultra-low delay requirements, we intend to design a joint allocation of local processor clock speed and uplink transmission power for each user under the constantly changing computation task load and wireless channel conditions. While solving the problem centrally at the CPU can allow for efficient joint resource allocation, global knowledge of the entire network state is required for decision making. For instance, information originating from users’ side such as computing demands, maximum tolerable task execution deadlines, and channel conditions should be collected, processed and then the allocation decisions need to be communicated back to the users within the tight delay tolerance, incurring prohibitively large overhead and additional delay for the two-way information exchange.  On the other hand, distributed approaches based on traditional optimization methods may lead to suboptimal performance due to lack of a central coordinator. Particularly, in a multi-user MEC system, where users compete for a limited communication and computing resources to accomplish computational task execution under stringent delay constraints, conceiving a globally optimal distributed conventional algorithm, that is also robust to time-varying computation task loads of the users, \textcolor{black}{solely based on local observations is a challenging task.} In general, applying conventional optimization methods or other heuristic algorithms to solve the JCCRA problem in a dynamic mobile edge network necessitates frequent re-evaluation of the optimal allocation strategy, following the non-deterministic task arrivals with time-varying task size and stochastic wireless channel conditions. It is also difficult to ensure a steady long-term performance with conventional algorithms relying on a one-shot optimization \textcolor{black}{[11]}. Furthermore, the algorithms must converge within the ultra-low delay tolerance. Consequently, the application of these methods to support time-critical and computation-intensive services is critically limited in practice. The JCCRA problem, therefore, calls for adaptive and robust solution approaches with reasonable complexity and signaling overheads.

Recently, integrating the capabilities of artificial intelligence (AI) in the mobile edge network, referred to as edge intelligence, is considered as one of the key enablers to realize the highly delay-intolerant advanced multimedia applications, e.g., extended reality (XR) \textcolor{black}{[12]}. In particular, owing to their adaptability in dynamic systems, reinforcement learning algorithms can be applied at different entities in the mobile edge network to derive efficient resource allocation strategies. Noting the above-mentioned difficulties associated with conventional optimization methods, as well as the additional delay and overhead for a two-way information exchange incurred in centralized JCCRA approach, we propose a distributed solution approach based on cooperative multi-agent reinforcement learning (MARL) for the formulated JCCRA problem. More specifically, each user is implemented as a learning agent that makes joint resource allocation relying on local observations only. The agents are trained with the state-of-the-art multi-agent deep deterministic policy gradient (MADDPG) algorithm under the framework of centralized training and decentralized execution. As each agent interacts continuously with the mobile edge network, it tries to capture the underlying dynamics of the environment and makes use of the acquired knowledge for an adaptive and robust joint resource allocation during the online execution phase. Our simulation results show that the performance of the proposed distributed approach outperforms the conventional baselines, while converging to that of a centralized deep deterministic policy gradient (DDPG)-based target benchmark, without the need for central processing and resorting to additional overhead and time cost. Furthermore, a significant performance improvement in terms of providing energy-efficient and consistently low latency task offloading has been demonstrated as compared to cellular MEC systems, namely small-cell network and co-located massive MIMO system. To the best of our knowledge, this is the very first attempt to solve JCCRA problem in a distributed fashion for cell-free network enabled mobile edge computing network. The intelligent distributed JCCRA scheme coupled with the reliable performance of the cell-free massive MIMO architecture in our framework can be a promising means of handling the stringent requirements of the envisaged multimedia applications, such as holographic displays, XR and others.

\begin{table}
	\caption{Summary of important notations}
	\centering
	\renewcommand{\arraystretch}{1.3}
	\setlength{\tabcolsep}{8pt} 
	\begin{tabular}{lp{0.6\linewidth}}
		\toprule
		\textbf{Notation} & \textbf{Description} \\
		\hline
		\midrule
		$\mathcal{M}=\left\{ 1,2,...,M \right\}$ & Set of access point (AP) indices \\
		$\mathcal{K}=\left\{ 1,2,...,K \right\}$ & Set of user indices  \\
		${{{\cal C}}_k}$  & Cluster of APs to serve user $k$ \\
		${N_k}$ & Number of APs in cluster for user $k$ \\
		$AP_{n}^{(k)}$  & $n$-th AP in ${{{\cal C}}_k}$ serving user $k$, $n=1,2,...,{{N}_{k}}$ \\
		$\Delta t$  & Duration of a time step \\
		$t_k^d$  & Application deadline for user $k$ \\
		${{\alpha }_{k}}$  & Proportion of local processor clock speed allocated to user $k$ \\
		${{\eta }_{k}}$  & Uplink transmit power control coefficient of user $k$ \\
		$p_{k}^{\max }$ / $p_{k}$ & Max / allocated uplink transmit power by user $k$ \\
		$W$ & System bandwidth \\
		${{R}_{k}}$  & Uplink rate of user $k$ \\
		${{N}_{cpb}}$  & Number of CPU cycles required to process one-bit task \\
		${{\cal T}_{k}}$\hspace*{0.26mm}/ ${\cal T}_k^{\rm{local}}$\hspace*{0.27mm}/ ${\cal T}_{k}^{\rm{offload}}$  & Incoming / locally computed / offloaded task size (in bits) by user $k$ \\
		$f_{k}^{\max }$ / $f_{k}^{\rm{local}}$ & Max / allocated local processor clock speed (in Hz) by user $k$ \\
		${{f^{\rm{CPU}}}}$  & Computing clock speed of the edge server in the CPU (in Hz) \\
		$f_{k}^{\rm{CPU}}$ & Allocated computational resource at the CPU for user $k$ \\
		$t_k^{tr}$ / $t_{k}^{comp}$ & Transmission / computing delay experienced by user $k$ \\
		$t_{k}^{\rm{local}}$ / $t_{k}^{\rm{offload}}$ / ${{t}_{k}}$ & Local execution / offloading / total delay experienced by user $k$ \\
		$E_{k}^{\rm{local}}$\hspace*{0.35mm}/$E_{k}^{\rm{offload}}$\hspace*{0.3mm}/ ${{E}_{k}}$ & Local / offloading / total energy consumption incurred by user $k$ \\
		\bottomrule
	\end{tabular}
\vspace{-0.35cm} 
\end{table}

The rest of the paper is organized as follows: In Section II, we review related works on JCCRA allocation in different MEC system models. In Section III, we discuss the system model for the considered mobile edge network and then, present problem formulation for JCCRA. Section IV discusses the proposed distributed cooperative multi-agent reinforcement learning-based solution approach for the JCCRA problem. Simulation results are presented in Section V, followed by some concluding remarks in Section VI.  For convenience, we summarized the key notations to be used throughout the paper in Table I. 

\section{Related Work}
\textcolor{black}{Recently, several centralized computation offloading and resource allocation schemes have been proposed for different MEC system models, targeting latency minimization, energy consumption reduction, or delay-energy tradeoff balancing. For instance, [13], and [14] analyze a joint computation offloading and resource allocation problem to minimize latency in vehicular networks, and in energy harvesting devices, respectively.} Minimization of energy consumption with partial offloading in a single-user MEC system is investigated in \textcolor{black}{[15]}, which was later extended to a multi-user MEC system by jointly optimizing computation and communication resources \textcolor{black}{[16]}. \textcolor{black}{Therein, dynamic voltage scaling (DVS) technique was adopted to adjust the frequency of the processor. Additionally, [17] studies energy-efficient resource allocation in latency constrained three-node MEC system, while the cost of cooling edge serve is included for wirelessly powered devices [18]. Meanwhile, energy-delay tradeoff is examined in [19] for a binary offloading MEC system, in which the users decide to offload or not. Similarly, power-delay tradeoff for partial offloading MEC system based on Lyapunov optimization is studied in [20].} We note that the above centralized schemes are based on conventional optimization methods that are typically limited to quasi-static systems, e.g., deterministic task models and time-invariant channel conditions, or require accurate knowledge of network wide information \textcolor{black}{[21]}, such as task size, application requirements, energy level of battery, and channel conditions of the users, which is difficult to obtain in practice under uncertain wireless network within the tight delay constraint. Therefore, their application to support time-critical and computation-intensive services might be hindered due to the associated communication and signaling overheads and additional time cost. Moreover, the centralized algorithms are not scalable in the sense that their computational complexity increases with system size. It is even more challenging from a cell-free massive MIMO point of view as we have to deal with large number of access points (APs) in a joint manner, in contrast to the above cell-centric MEC models, where a user is connected to only single AP in each cell.

To relieve the signaling overhead and address scalability issues of the centralized resource allocation schemes, \textcolor{black}{a} plethora of distributed approaches have been proposed. A design of distributed offloading decision is formulated as a multi-user offloading game among the users for mobile cloud computing \textcolor{black}{[22]}, wireless-powered MEC networks \textcolor{black}{[23]}, and MEC-empowered small-cell networks \textcolor{black}{[24]}. In \textcolor{black}{[25]}, offloading decision and resource allocation problem is solved sequentially through decomposition technique within proximate clouds in a distributed fashion. However, the focus of the majority distributed JCCRA approaches is confined to the design of offloading decisions. In addition, due to a lack of a central coordinator, the algorithms may converge to suboptimal solutions. It is worth mentioning that all these schemes are based on conventional optimization methods that are limited to quasi-static systems with deterministic task arrival and invariant channel conditions, failing to capture the randomness in practical systems. Consequently, in response to the dynamics in the environment, the JCCRA problem should be solved frequently, and the algorithms are expected to converge swiftly within the task execution deadline. Such approaches are, therefore, not suited for handling time-sensitive applications.

JCCRA in a dynamic mobile edge computing network with time-varying channel conditions and stochastic task arrivals require robust solutions that can not only adapt to the diversities in the network conditions and application-specific requirements (e.g., delay constraint), but also perform well with constrained radio and computational resources. In this regard, owing to their adaptability in dynamic settings, reinforcement learning-based frameworks have recently been proposed to solve joint resource allocation problems in various MEC systems. For instance, a Q-learning based design of offloading decisions has been proposed for a single-cell MEC \textcolor{black}{system [28]}, while a deep Q-network (DQN) algorithm has been proposed to solve binary offloading decisions in different MEC system models in \textcolor{black}{[27], and [28]. Furthermore, [29], and [30] utilize a deep deterministic policy gradient (DDPG) algorithm for joint resource allocation schemes in multi-server MEC systems.} In all of these works, however, the decisions are made by central entities which need to gather system-wide information; thus, the schemes are prone to the shortcomings of centralized resource allocations discussed above. Moreover, training a centralized super-agent can potentially be ineffective as the convergence of the training is not guaranteed.  \textcolor{black}{[33]} investigates a distributed JCCRA problem in a single-cell MEC system in which each user tries to minimize the individual cost, defined as a weighted sum of power and delay, in a non-cooperative fashion, possibly resulting in suboptimal performance. Moreover, the adopted computational model is not applicable to handle computation-intensive tasks, requiring extreme reliability and ultra-low hard deadline. 

Recently, there are few works on the cell-free massive MIMO-based MEC systems. In a MEC system with both CPU and APs equipped with computing servers, several performance analyses are presented in \textcolor{black}{[32]}, while considering coverage radius of the APs and probabilities of offloading to the servers. From a massive access perspective, the issues of active user detection and channel estimations are discussed for a cell-free massive MIMO-based computing framework in \textcolor{black}{[33]}. In contrast to our work, both \textcolor{black}{[34] and [35]} do not deal with the JCCRA problem. A centralized resource allocation scheme based on an iterative conventional optimization algorithm is proposed in \textcolor{black}{[34]}. In a dynamic MEC system, where optimal solutions are recalculated frequently, the algorithm may become computationally expensive. \textcolor{black}{From the perspective of improving energy efficiency, [35] highlights the potential of integrating reconfigurable intelligent surface (RIS)-aided cell-free massive MIMO system with wireless energy transfer (WET) to extend the life-span of devices with limited battery power. Meanwhile, [36] explores the optimization of transmit power for access points (APs), and temporary deactivation of a subset of APs in cell-free network to maximize energy efficiency, while meeting users’ spectral efficiency requirements. }
\section{System Model and Problem Formulation}
\subsection{User-centric Cell-free Massive MIMO: Overview}
A cell-free massive MIMO system is considered as a potential beyond-5G and 6G network infrastructure as it provides a uniform user experience by eliminating the cell edge. It exploits joint signal processing from a large number of distributed access points (APs), which simultaneously serve a much smaller number of users using the same radio resources. To attain the tight requirements of the envisaged advanced services, the capability of cell-free access links to provide reliable performance without cell edge is essential to play the key functionality of a virtual bus while offloading an intensive task for parallel computation at the edge server. 

We consider a user-centric mobile edge network with $M$ single-antenna access points (APs) and $K$  single-antenna users, where $M\gg K$ , entailing to the widely known model for cell-free massive MIMO system \textcolor{black}{[9]}. The APs are all connected to a central processing unit (CPU) via error-free backhaul links. Let  $\mathcal{M}=\left\{ 1,2,...,M \right\}$ and  $\mathcal{K}=\left\{ 1,2,...,K \right\}$ denote sets of AP and user indices, respectively. We assume that all APs and UEs are distributed uniformly throughout the network. Fig. 1 illustrates a system model for cell-free massive MIMO-enabled mobile edge network. Therein, the CPU is equipped with a computing server of finite computational capability to provide computing resources for users with computationally intensive and latency stringent applications. Each user is also equipped with limited local processing capability.

\begin{figure}[!t]       
\centering
\includegraphics[width=2.6in,height=2.3in]{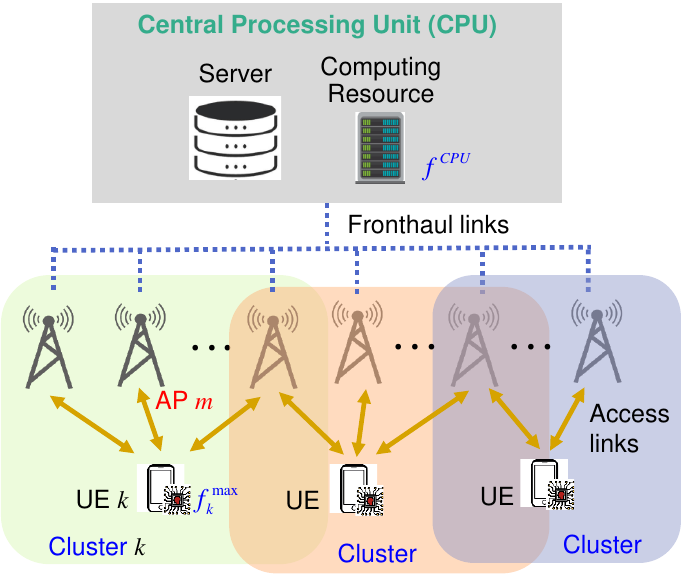}

\caption{User-centric cell-free massive MIMO-enabled mobile edge network: \emph{Illustrative system model}}
\vspace{-10pt}
\label{fig_sim}
\end{figure}

Let the channel between $m$-th AP and $k$-th user is given as ${g_{mk}}=\beta _{mk}^{{}^{1}/{}_{2}}{h_{mk}}$ where ${{\beta }_{mk}}$ is a large-scale channel gain coefficient and \textcolor{black}{${h_{mk}} \sim {{\cal C}{\cal N}}\left( {0,1} \right)$} represents small-scale channel fading. Let ${{\tau }_{c}}$ denote a channel coherence period in which ${{h}_{mk}}$ remains the same, while ${{\beta }_{mk}}$ remains constant for a time interval which lapses multiple channel coherence times. The coherence period ${{\tau }_{c}}$ is divided into pilot transmission time of ${{\tau }_{p}}$ samples and uplink data transmission time of (${{\tau }_{c}}-{{\tau }_{p}}$) samples.

All users simultaneously transmit pilot sequences to the APs for uplink channel estimation. Let ${{\mathbf{\psi }}_{k}}\in {{\mathsf{\mathbb{C}}}^{{{\tau }_{p}}\times \,1}}$ denote a pilot sequence of user $k$ where  ${{\left\| {{\mathbf{\psi }}_{k}} \right\|}^{2}}=1$. We assume that pilot contamination can be ignored by picking pairwise orthogonal sequences {${{\mathbf{\psi }}_{1}},{{\mathbf{\psi }}_{2}},.....,{{\mathbf{\psi }}_{K}}$}, i.e., ${{\tau }_{p}}=K$.   Then, the received pilot vector at the $m$-th AP, $\mathbf{y}_{m}^{p}\in {{\mathsf{\mathbb{C}}}^{{{\tau }_{p}}\times \,1}}$ , can be represented as follows:
\begin{equation}
{\bf{y}}_m^p = \sqrt {{\tau _p}} \sum\limits_{k = 1}^K {\sqrt {p_k^p} {g_{mk}}{{\bf{\psi }}_k}}  + {\bf{\omega }}_m^p,                              
\end{equation}
where $p_{k}^{p}$ denotes the pilot transmit power, and $\mathbf{\omega }_{m}^{p}$ denotes a $\,{{\tau }_{p}}$-dimensional additive noise vector with independent and identically distributed entries of \,${\bf{\omega }}_m^p \sim {{\cal C}{\cal N}}\left( {0,\sigma _{m}^{2}} \right)$. Based on the received vector ${\bf{y}}_m^p$, the least-square (LS) channel estimate ${\hat g_{mk}}$  can be expressed as \textcolor{black}{[10]}
\begin{equation}
{\hat g_{mk}} = \frac{1}{{\sqrt {{\tau _p}p_k^p} }}{\bf{\psi }}_k^H{\bf{y}}_m^p.                      
\end{equation}
The channel estimates are used to decode the uplink transmitted data of the users. After pilot transmission, the users transmit offloaded data to the APs. Let ${{x}_{k}}$ denote the uplink transmission data of user $k$. Then, the received signal $y_{m}^{u}$ at the $m$-th AP is given as
\begin{equation}
y_m^u = \sum\limits_{k = 1}^K {\sqrt {{p_k}} {g_{mk}}{x_k}}  + {\omega _m},                                              
\end{equation}
where ${{p}_{k}}$ is the uplink data transmit power. Let ${{p}_{k}}={{\eta }_{k}}\,p_{k}^{\max }$, where ${{\eta }_{k}}$ and $p_{k}^{\max }$ represent \textcolor{black}{the} power control coefficient and maximum uplink transmit power of UE $k$, respectively. 

In order to ensure scalability of cell-free massive MIMO system in terms of complexity for pilot detection and data processing, each UE is served by a limited number of APs, which forms a cluster. To this end, it is essential to form a cluster of APs serving each user in a user-centric manner. Limiting the total number of APs serving the $k$-th user to ${{N}_{k}}$, such that  ${{N}_{k}}\leq M$, we then form a user-centric cluster of APs  ${{{\cal C}}_k} = \left\{ {AP_1^{(k)},AP_2^{(k)}, \cdots ,AP_{{N_k}}^{(k)}} \right\}$, where $AP_{n}^{(k)}$ \textcolor{black}{denotes} the $n$-th AP in the cluster. For simplicity of exposition in the current discussion, assume that all users have the same size of cluster, given by ${{C}^{\max }}$. In order to construct a cluster ${{{\cal C}}_k}$, we employ a greedy approach which orders the large-scale channel coefficients ${\beta _{mk}}$ of user $k$ with the respective APs in a descending order and then, includes all APs with the largest ${\beta _{mk}}$ until  ${{N}_{k}}={{C}^{\max }},\text{ }k=1,2,\cdot \cdot \cdot ,K$.  Thus, the transmitted data by user $k$ is only decoded by the APs in ${\cal C}_k$. Then, each AP $m\in {\cal C}_k$ transmits the quantity $\hat{g}_{mk}^{*}y_{m}^{u}$ to the CPU via a fronthaul link. The received soft estimates are combined at the CPU to decode the data transmitted by user $k$ as follows:  
\begin{equation}
{\hat x_k} = \sum\limits_{m = 1}^{N_k} {\hat g_{mk}^ * y_m^u}.
\end{equation}
Then, the uplink SINR ${{\gamma }_{k}}$ for user $k$ can be expressed as
\begin{equation}
{\gamma _k} = \frac{{{p_k}{{\left| {\sum\limits_{m \in {{{\cal C}}_k}} {\hat g_{mk}^ * {g_{mk}}} } \right|}^2}}}{{\sum\limits_{k' \ne k} {{p_{k'}}} {{\left| {\sum\limits_{m \in {{{\cal C}}_k}} {\hat g_{mk}^ * {g_{mk'}}} } \right|}^2} + \sigma _m^2{{\left| {\sum\limits_{m \in {{{\cal C}}_k}} {{{\hat g}_{mk}}} } \right|}^2}}}\,\,.
\end{equation}    
The uplink rate of user $k$ is then given as \textcolor{black}{${R_k} = W\left( {{{\left( {{\tau _c} - {\tau _p}} \right)} \mathord{\left/
 {\vphantom {{\left( {{\tau _c} - {\tau _p}} \right)} {{\tau _c}}}} \right.
 \kern-\nulldelimiterspace} {{\tau _c}}}} \right){\log _2}\left( {1 + {\gamma _k}} \right)$}, where $W$ is the system bandwidth. 
\subsection{Parallel Computation Model}
Without loss of generality, we assume each user $k$ has a computationally intensive task with ${{{\cal T}}_{k}}\left( t \right)$ bits at the beginning of every discrete time step  $t = 1,2,...$, whose duration is set to a coherence period, i.e., $\Delta t={{\tau }_{c}}$. The task sizes in different time steps are independent and uniformly distributed over $\left[ {{{\cal T}}_{\min }},{{{\cal T}}_{\max }} \right]$, for every user $k\in {{\cal K}}$. Let $t_{k}^{d}$ denote a strict deadline of the $k$-th user to complete execution of the time-sensitive application, e.g., tactile internet \textcolor{black}{[6]}. Furthermore, we assume the tasks are independent and fine grained, i.e., can be broken into arbitrary portions so that they can be computed at the user device locally and the edge server in parallel, similar to \textcolor{black}{[15] and [16]}. At time step $t$, considering the delay constraint and energy consumption, the $k$-th user processes ${{\cal T}}_{k}^{\textcolor{black}{\rm{local}}}\left( t \right)$ bits locally and offloads the remaining ${{\cal T}}_{k}^{\textcolor{black}{\rm{offload}}}\left( t \right)$ bits to the edge server at the CPU. In the sequel, we describe the models adopted for local computation and computation offloading.

\subsubsection{Local Computation} 
Let us denote the maximum local computing clock speed of user $k$  by $f_{k}^{\max }$ (in cycle per second). Furthermore, let ${{N}_{cpb}}$ denote the number of CPU cycles to process a one-bit task. Taking energy consumption and delay requirement of the application into account, the user decides the proportion ${{\alpha }_{k}} \in\left[ 0,1 \right]$ for achieving the local clock speed of $f_k^{\textcolor{black}{{\rm{local}}}}\left( t \right)={{\alpha }_{k}}\left( t \right)f_{k}^{\max }$, which in turn determines the size of locally computed task. The entire task ${{{\cal T}}_{k}}\left( t \right)$ is offloaded to the edge server if $\,{\alpha _k}\left( t \right) = 0$. Meanwhile, ${\alpha _k}\left( t \right) = 1$ if the whole local processing capability of $f_{k}^{\max }$ is fully utilized while offloading the remaining task bits to the edge server. Note that conservative local processing may lead to high energy consumption, while aggressive computation offloading by all users may subject some users to service outage, since offloading all tasks cannot be supported with finite radio and computation resources. In general, ${\alpha _k}$ is one of the decision variables for efficient JCCRA, which is governed by the channel and computational constraints in the local and edge systems, subject to the energy consumption and delay constraints. 

Given the application deadline $t_{k}^{d}$ to process ${{{\cal T}}_{k}}\left( t \right)$ task bits subject to the local processing clock speed of $f_{k}^{\rm{local}}\left( t \right)$, then the size of locally computed task can be expressed as ${{\cal T}}_{k}^{\rm{local}}\left( t \right)=\min \left( {{{\cal T}}_{k}}\left( t \right),\frac{t_{k}^{d}f_{k}^{\rm{local}}\left( t \right)}{{{N}_{cpb}}} \right)$ (in bits).  Let $t_{k}^{\textcolor{black}{\rm{local}}}\left( t \right)$ denote the time taken for local execution at time step $t$, which is given as
\begin{equation}
t_k^{\rm{local}}\left( t \right) = \min \left( {\frac{{{{\cal T}}_k^{\rm{local}}\left( t \right){N_{cpb}}}}{{f_k^{\rm{local}}\left( t \right)}},t_k^d} \right).
\end{equation}
Then, the energy consumed for the local execution is expressed as
\begin{equation}
E_k^{\textcolor{black}{\rm{local}}}\left( t \right) = \varsigma\, {{\cal T}}_k^{\rm{local}}\left( t \right){N_{cpb}}{\left( {f_k^{\rm{local}}\left( t \right)} \right)^2},
\end{equation}
where $\varsigma$ corresponds to the effective switched capacitance depending on the chip architecture. 
\subsubsection{Computation Offloading}
The $k$-th user offloads the remaining task bits ${{\cal T}}_{k}^{\textcolor{black}{\rm{offload}}}\left( t \right)=\max \left( 0,\,\,{{{\cal T}}_{k}}\left( t \right)-{{\cal T}}_{k}^{\rm{local}}\left( t \right) \right)$ to the edge server at the CPU for parallel computation. While offloading the computational task to the edge server, the experienced latency can be broken down into transmission delay for \textcolor{black}{offloading} data, processing delay in edge server, and transmission delay for retrieving the result. Since the retrieved data size after computation in the server is much smaller as compared to the offloaded data size, we ignore the retrieving delay in our formulation. Let $t_{k}^{tr}\left( t \right)$ and $t_{k}^{comp}\left( t \right)$ denote the transmission delay and computing delay, respectively, for the $k$-th user. To offload ${{\cal T}}_{k}^{\rm{offload}}\left( t \right)$ data bits to the edge server, the transmission delay $t_{k}^{tr}\left( t \right)$  is given as   
\begin{equation}
t_k^{tr}\left( t \right) = \frac{{{{\cal T}}_k^{\rm{offload}}\left( t \right)}}{{{R_k}({\eta _k},t)}},
\end{equation}
where ${R_k}(t)$ is the uplink rate of user $k$ at time step $t$, which is expressed according to the discussion in the previous subsection. 

Let ${{f}^{\rm{CPU}}}$ (in cycle per second) denote the computing clock speed of the edge server in the  CPU which is shared among the users in proportion to the offloaded task size so that each user can experience uniform computation delay. In other words, the allocated computational resource at the CPU for user $k$, denoted as ${{f}_{k}}^{\rm{CPU}}\left( t \right)$,  can be expressed as 
\begin{equation}
f_k^{\rm{CPU}}\left( t \right) = \frac{{{{\cal T}}_k^{\rm{offload}}\left( t \right)}}{{\sum\limits_{k = 1}^K {{{\cal T}}_k^{\rm{offload}}\left( t \right)} }}{f^{\rm{CPU}}}.
\end{equation}
Then the computing time $t_{k}^{comp}\left( t \right)$ required to execute  bits ${{\cal T}}_{k}^{\rm{offload}}\left( t \right)$ is given as
\begin{equation}
t_k^{comp}\left( t \right) = \frac{{{{\cal T}}_k^{\rm{offload}}\left( t \right){N_{cpb}}}}{{f_k^{\rm{CPU}}\left( t \right)}}.
\end{equation}
Therefore, the total edge-computing delay for executing ${{\cal T}}_{k}^{\rm{offload}}\left( t \right)$ bits is given as the sum of the delays for uplink transmission and computation at the CPU, i.e., $t_{k}^{\textcolor{black}{\rm{offload}}}\left( t \right)=t_{k}^{comp}\left( t \right)+t_{k}^{tr}\left( t \right)$. The corresponding energy consumption is given as
\begin{equation}
E_k^{{\textcolor{black}{\rm{offload}}}}\left( t \right) = {p_k}\left( t \right)t_k^{tr}\left( t \right),
\end{equation}
where the transmission power at time step $t$, ${{p}_{k}}\left( t \right)$, is expressed as ${{p}_{k}}\left( t \right)={{\eta }_{k}}\,\left( t \right)p_{k}^{\max }$, where $p_{k}^{\max }$ and ${\eta _k}\,\left( t \right)$ correspond to the maximum uplink transmission power and power control factor of UE $k$, respectively. It should be noted that power control plays a critical role as it governs co-channel interference among the different clusters during computation offloading, in addition to determining the energy consumption for offloading. To that end, ${\eta _k}$  is another variable to consider for efficient utilization of limited radio and computing resources. 

 According to the communication and parallel computation models discussed above, the overall experienced latency ${{t}_{k}}\left( t \right)$ by user $k$, to execute  ${{{\cal T}}_{k}}\left( t \right)$ bits locally and at the edge is given by
\begin{equation}
{t_k}\left( t \right) = \max (t_k^{\rm{local}}\left( t \right),t_k^{\rm{offload}}\left( t \right)).
\end{equation}
The corresponding energy consumption ${{E}_{k}}\left( t \right)$ of user $k$, can be expressed as 
\begin{equation}
{E_k}\left( t \right) = E_k^{\rm{local}}\left( t \right) + E_k^{\rm{offload}}\left( t \right).
\end{equation}
\subsection{JCCRA Problem Formulation}
We now present JCCRA problem formulation in our framework. Specifically, our objective is to minimize the total energy consumption of all users while meeting the respective user-specific delay requirements by jointly optimizing the local processor speed $f_{k}^{\rm{local}}\left( t \right)={{\alpha }_{k}}\left( t \right)f_{k}^{\max }$, and uplink transmission power ${{p}_{k}}\left( t \right)={{\eta }_{k}}\,\left( t \right)p_{k}^{\max }$, for every user $k \in {{\cal K}}$ at each time step $t$, without any prior assumptions on the realizations of the incoming task size and channel conditions of the users. The JCCRA problem to jointly determine $\left( {{\alpha }_{k}}\left( t \right),{{\eta }_{k}}\left( t \right) \right),\forall k\in {{\cal K}}$ can mathematically be formulated by (14), as shown at the top of this page. Therein, the first constraint ensures that the task execution delay ${{t}_{k}}\left( t \right)$ should not exceed the user-specific delay requirement $t_{k}^{d}$. 

\begin{table*}[ht]
\begin{equation}
	\begin{array}{l}
		\,\mathop {\min }\limits_{\{ {\alpha _k}(t),{\eta _k}(t)|\forall k\} } \,\,\,\,\,\,\,\,\,\sum\limits_{k = 1}^K {{E_k}(t) = \sum\limits_{k = 1}^K {\varsigma \min \left( {{{{\cal T}}_k}(t),\frac{{t_k^d{\alpha _k}(t)f_k^{\max }}}{{{N_{cpb}}}}} \right){N_{cpb}}{{\left( {{\alpha _k}(t)f_k^{\max }} \right)}^2} + {\eta _k}(t)\,p_k^{\max }\frac{{\max \left( {0,\,\,{{{\cal T}}_k}(t) - \min \left( {{{{\cal T}}_k}(t),\frac{{t_k^d\,{\alpha _k}(t)f_k^{\max }}}{{{N_{cpb}}}}} \right)} \right)}}{{{{R_k}(t)}}}} } \\
		\,\,\,\,\,\,\,\,\,\,{\rm{subject}}\,{\rm{to}}\,\,\,\,\,\,\,\,\,\,\,\,\,\,\max \left( {\min \left( {\frac{{\min \left( {{{{\cal T}}_k}(t),\frac{{t_k^d\,{\alpha _k}(t)f_k^{\max }}}{{{N_{cpb}}}}} \right){N_{cpb}}}}{{{\alpha _k}(t)f_k^{\max }}},t_k^d} \right),{{\cal T}}_k^{\rm{offload}}\left( t \right)\left[ {\frac{{{N_{cpb}}}}{{f_k^{\rm{CPU}}\left( t \right)}} + \frac{1}{{{R_k}(t)}}} \right]} \right) \le t_k^d,\,\,\forall k\\
		\,\,\,\,\,\,\,\,\,\,\,\,\,\,\,\,\,\,\,\,\,\,\,\,\,\,\,\,\,\,\,\,\,\,\,\,\,\,\,\,\,\,\,\,\,\,\,\,\,\,\,\,\,0 \le {\alpha _k}\left( t \right) \le 1,\,\,\forall k\\
		\,\,\,\,\,\,\,\,\,\,\,\,\,\,\,\,\,\,\,\,\,\,\,\,\,\,\,\,\,\,\,\,\,\,\,\,\,\,\,\,\,\,\,\,\,\,\,\,\,\,\,\,\,0 \le {\eta _k}\left( t \right) \le 1,\,\,\,\forall k
	\end{array}
\end{equation}
\noindent\textbf{\rule[0.5ex]{\linewidth}{1.00pt}}

\end{table*}

The JCCRA problem in (14) is a stochastic optimization problem in which the objective function and the first condition involve constantly changing random variables, following the dynamics caused by the random incoming task size at each user, and wireless channel conditions. Hence, the time-varying optimization variables  $\left( {{\alpha }_{k}}\left( t \right),{{\eta }_{k}}\left( t \right) \right),\forall k\in {{\cal K}}$, should be determined frequently, i.e., at every step $t$, and rapidly within the ultra-low deadline. It is, therefore, challenging to efficiently solve the problem using conventional optimization algorithms in an adaptive manner with a reasonable computational complexity. \textcolor{black}{Moreover, the distributed nature of the cell-free massive MIMO systems introduces unique challenges for JCCRA. Unlike the traditional cellular systems, where base stations have fixed coverage areas with limited number of users, cell-free systems have multiple access points (APs) serving users, with typically no fixed number of users per AP, leading to more complex and dynamic interference patterns, which must be managed effectively for optimal joint resource allocation. It is also important to mention that solving the JCCRA problem centrally at the CPU} might allow for efficient management of the available radio and computing resources due to globally processing network-wide information, however, the associated overheads and additional delay, i.e., time cost, for the two-way information exchange are forbiddingly significant, especially given the ultra-low delay constraints. Thus, to cope up with the dynamics in the mobile edge network and derive a flexible and efficient joint resource allocation for every user, we propose a distributed JCCRA based on cooperative multi-agent reinforcement learning. We note that this is the very first attempt to solve JCCRA problem in a distributed manner for a cell-free enabled mobile edge computing network. While alleviating the overheads and time cost associated with the centralized implementation, the proposed distributed approach enables users to entertain energy-efficient and consistently low end-to-end delay computational task offloading.

\section{The Proposed Multi-Agent Reinforcement Learning-based Distributed JCCRA}
In this section, we present a novel distributed JCCRA solution approach based on cooperative multi-agent reinforcement learning framework, in which each user device is implemented as an agent for joint resource allocation relying on local observation only. In particular, the agents learn to map their local observation into JCCRA actions that minimize the total energy consumption while meeting the respective application deadlines.  
\begin{figure}[t]
\centering
\includegraphics[width=3.4in,height=2.5in]{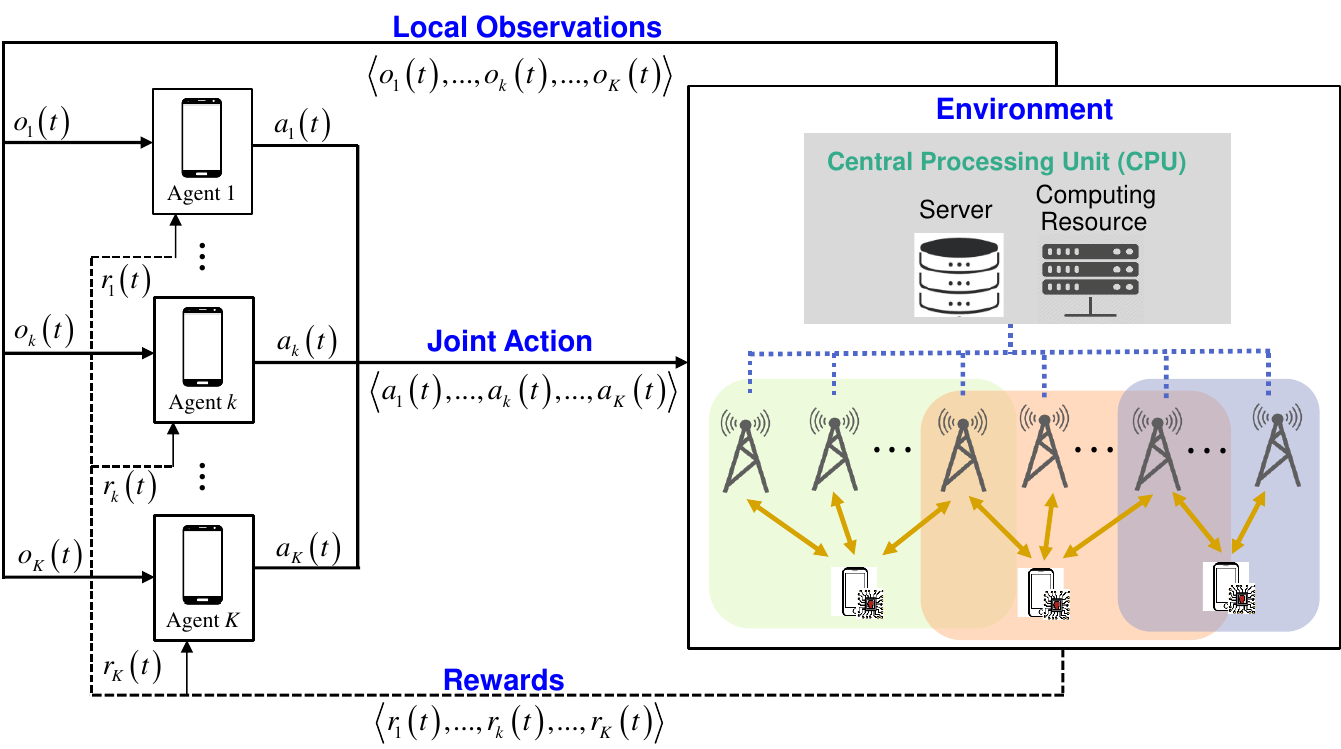}

\caption{Multi-agent reinforcement learning framework: \emph{Illustration}}
\label{fig_sim}
\end{figure}

\subsection{Distributed DRL Formulation for JCCRA}
The DRL agents sequentially interact with the mobile edge network in discrete-time steps to learn optimal joint resource allocation policies. Let ${{{\cal O}}_k}$, and ${{\cal S}}$ denote the local observation space of agent $k \in {{\cal K}}$, and the complete environment state space, respectively. As shown in Fig. 2, at time step $t$, each agent relies on local observation of the environment state ${o_k}\left( t \right):{{\cal S}} \mapsto {{{\cal O}}_k}$ to determine an action ${a_k}\left( t \right) \in {{{\cal A}}_k}$, from its action space ${ {{\cal A}}_k}$ according to the current JCCRA policy ${{\mu }_{k}}$. The shared environment collects the joint action of the agents $a\left( t \right)=\left( {{a}_{1}}\left( t \right),...,{{a}_{K}}\left( t \right) \right)$, and emits the next observations ${o_k}\left( {t + 1} \right) \in { {{\cal O}}_k}$ and real-valued scalar rewards ${r_k}\left( t \right):{{\cal S}} \times {{{\cal A}}_k} \mapsto \mathsf{\mathbb{R}}$ for all $k\in  {{\cal K}}$. The goal of the agents is, therefore, to constantly improve their respective policy until it converges to the optimal JCCRA policy $\mu _{k}^{*}$ that maximizes the expected long-term discounted cumulative reward, defined as ${{J}_{k}}\left( {{\mu }_{k}} \right)=\mathbb{E}\left[ \sum\limits_{t=1}^{T}{{{\varepsilon }^{t-1}}{{r}_{k}}}\left( t \right) \right]$, where $\varepsilon \in \left[ 0,\,1 \right]$ is the discount factor and  $T$ is the total number of total time steps (horizon). The optimal JCCRA policy  $\mu _{k}^{*}$ is then given as
\begin{equation}
\mu _k^ *  = \mathop {\arg \,}\limits_{{\mu _k}} \max {J_k}\left( {{\mu _k}} \right)\,.
\end{equation}

In the sequel, we define the \emph{local observation}, \emph{action}, and \emph{reward} of each agent $k \in {{\cal K}}$ for JCRRA at a given time $t$.
\subsubsection{Local observation}
As discussed in the previous section, the compute-intensive tasks are subjected to user-specific application deadline. Specifically, at time step $t$, the maximum delay tolerance of  the $k$-th agent to execute incoming task of ${{{\cal T}}_{k}}(t)$ bits is given by $t_{k}^{d}$. Meanwhile, we assume the agent has a rate assignment result from the previous time step, ${{R}_{k}}(t-1)$. At the beginning of each time step $t$, the local observation of agent $k$ is defined as
\begin{equation}
{o_k}\left( t \right) \buildrel \Delta \over = \left[ {{{{\cal T}}_k}(t),t_k^d,{R_k}(t - 1)} \right].
\end{equation}
\textcolor{black}{Here, ${{R}_{k}}(t-1)$, which is a function of the channel state information (CSI), serves as a proxy for the CSI in ${{o}_{k}}(t)$ to abstract the changes in wireless channel conditions. Please note that directly incorporating the channel estimates between the APs and the $k$-th user, i.e., ${\hat g_{mk}},\forall m \in {{\cal M}}$, in the local observation may not be practical due to the required high dimensional feedback, simply because the number of APs is typically much larger than the number of users, i.e. $M \gg K$, in cell-free massive MIMO system.}
\subsubsection{Action}
Based on the local observation of the environment, each agent $k$ jointly determines how much local computing resources and uplink transmit power must be allocated to execute ${{{\cal T}}_{k}}(t)$ within $t_{k}^{d}$. Accordingly, the action of the $k$-th agent at time step $t$ can be expressed as
\begin{equation}
{a_k}\left( t \right) \buildrel \Delta \over = \left[ {{\alpha _k}\left( t \right),{\eta _k}\left( t \right)} \right],
\end{equation}
where ${\alpha _k}(t) \in \left[ {0,1} \right]$, and ${\eta _k}(t) \in \left[ {0,1} \right]$ are continuous values that govern the local processor clock speed $f_{k}^{\rm{local}}\left( t \right)={{\alpha }_{k}}\left( t \right)f_{k}^{\max }$, and uplink transmit power ${{p}_{k}}={{\eta }_{k}}\,p_{k}^{\max }$, respectively.  
\subsubsection{Reward}
The agents must learn cooperative JCCRA policies that minimize the total energy consumption of all users while meeting the respective delay constraints. The immediate reward, therefore, should encapsulate both aspects of the design objective. Accordingly, to encourage the agents to maximize the common goal and enforce cooperation among them, we define the joint reward as 
\begin{equation}
{r_k}(t) =  - \sum\limits_{k = 1}^K {{\xi _k}{E_k}\left( t \right),\,\,} \forall k \in {{\cal K}},
\end{equation} 
where ${\xi _k} = 1$ if ${t_k}\left( t \right) \le t_k^d$, otherwise ${\xi _k} = 10$ to punish potentially selfish behavior of the agents that lead to failure in meeting the delay constraint. 
\begin{figure}[!t]
\centering

\includegraphics[width=3.4in,height=2.5in]{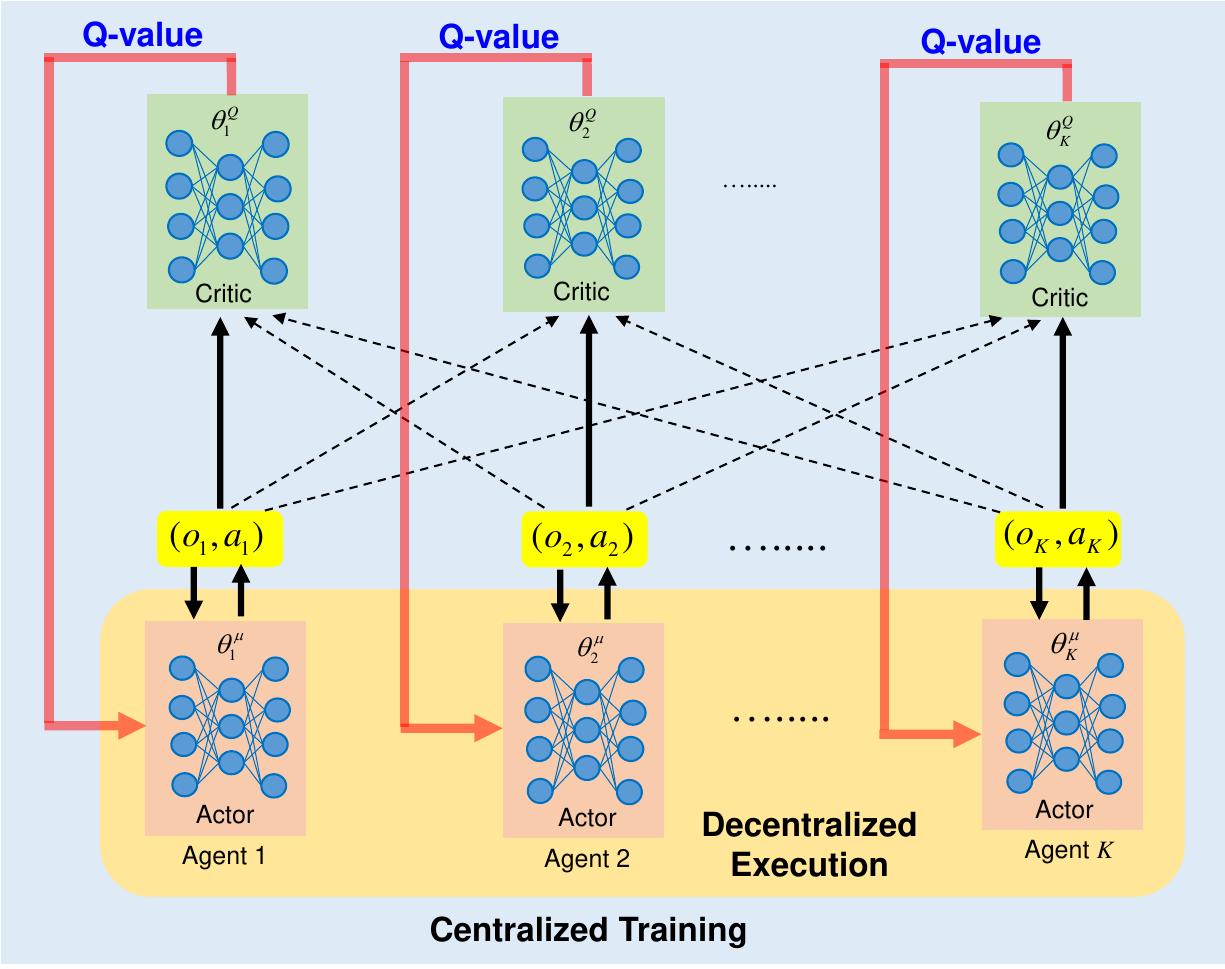}
\caption{Centralized training and decentralized execution framework in MADDPG algorithm}
\end{figure} 

\textcolor{black}{We note that the decision variables, i.e., ${\alpha _k}$  and ${\eta _k}$, for the JCCRA, are continuous-valued. One possible approach is to quantize the continuous action space into a finite set of actions and employ DRL algorithms designed for discrete action spaces, such as DQN [37]. Nevertheless, discretization into high-dimensional action space can cause the issue of the curse of dimensionality. Limited quantization levels, on the other hand, may lead to suboptimal policies due to loss of information from the action space (loss of granularity). Instead, reinforcement learning methods for continuous action spaces, such as deep deterministic policy gradient (DDPG) [38], can efficiently handle the curse of dimensionality in a continuous domain by employing actor-critic architecture, deep neural network-based function approximators for policy and value estimation, and learn compact representations of the state and action space, without explicitly enumerating them. However, a direct extension of single-agent DDPG algorithm, i.e., training each agent independently,}  faces several problems when applied to learn coordinated policies. Learning in a multi-agent setup is more challenging and complex than in single-agent case, since the environment is no longer stationary from the agent’s perspective as the other agents update their policies concurrently. In other words, the agents face a moving target problem, which may lead to learning instability. Moreover, the non-stationarity of the environment compounded on agents’ partial observations of the dynamic environment can cause shadowed equilibria, a phenomenon in which agents’ local optimal actions result in globally sub-optimal joint action [39]. In the next subsection, we discuss a multi-agent version of DDPG algorithm, referred to as multi-agent deep deterministic policy gradient (MADDPG) [40], based on the centralized training and decentralized execution framework, to train all agents for JCCRA decision making.

\begin{algorithm}[t]    
\DontPrintSemicolon

\For{each agent $k \in {{\cal K}}$}
{
	{Initialize replay buffer ${{{\cal D}}_k}$.}\;
	{Initialize the actor network ${\mu _k}\left( {{o_k}|\theta _k^\mu } \right)$ and critic network ${Q_k}\left( {{s_k},a|\theta _k^Q} \right)$ with weights $\theta _k^\mu $ and $\theta _k^Q$, respectively.}\;
	{Initialize the target networks, ${\mu '_k}$ and ${Q'_k}$ with weights $\theta _k^{\mu '}$ and 
		$\theta _k^{Q'}$, respectively.}\;
	
}\label{endfor}

\For{each episode $e = 1,2,...$}
{
	\For{each agent $k \in {{\cal K}}$}
	{  
		{Initialize random process ${{{\cal N}}_k}$ for exploration.}\;
		{Generate initial local observation from the environment simulator.}
		
	}\label{endfor}
	
	\For{each step $t = 1,2,...$}
	{  
		\For{each agent $k \in {{\cal K}}$}
		{  
			{Select action ${a_k}\left( t \right) = {\mu _k}\left( {{o_k}\left( t \right)|\theta _k^\mu } \right) + {{{\cal N}}_k}\left( t \right)$}
			
		}\label{endfor}
		Execute joint action $a\left( t \right) = \left( {{a_1}\left( t \right),....,{a_K}\left( t \right)} \right)$
		\For{each agent $k \in {{\cal K}}$}
		{  
			
			{Collect reward  ${r_k}\left( t \right)$ and observe ${s_k}\left( {t + 1} \right)$.}\;
			{Store the transition $\left( {{s_k}\left( t \right),a\left( t \right),{r_k}\left( t \right),{s_k}\left( {t + 1} \right)} \right)$ into ${{{\cal D}}_k}$.}\;
			{Sample random minibatch of \emph{B} transitions 
				$\left( {s_k^i,{a^i},r_k^i,s_k^{i + 1}} \right)$ from  ${{{\cal D}}_k}$.}\;
			{Update the critic network by minimizing the loss given by  (19).}\;
			{Update the actor policy using the sampled policy gradient given by (20).}\;
			{Update the target networks according to (21).}\;
			
		}\label{endfor}

	}\label{endfor}
}\label{endfor}

\caption{MADDPG Algorithm for JCCRA}\label{costalgorithm}
\end{algorithm}
\vspace*{-10pt}
\subsection{MADDPG Algorithm  for Distributed JCCRA}
\begin{figure*}[t]                                  
\centering
\includegraphics[width=\textwidth, height=3.5in, keepaspectratio]{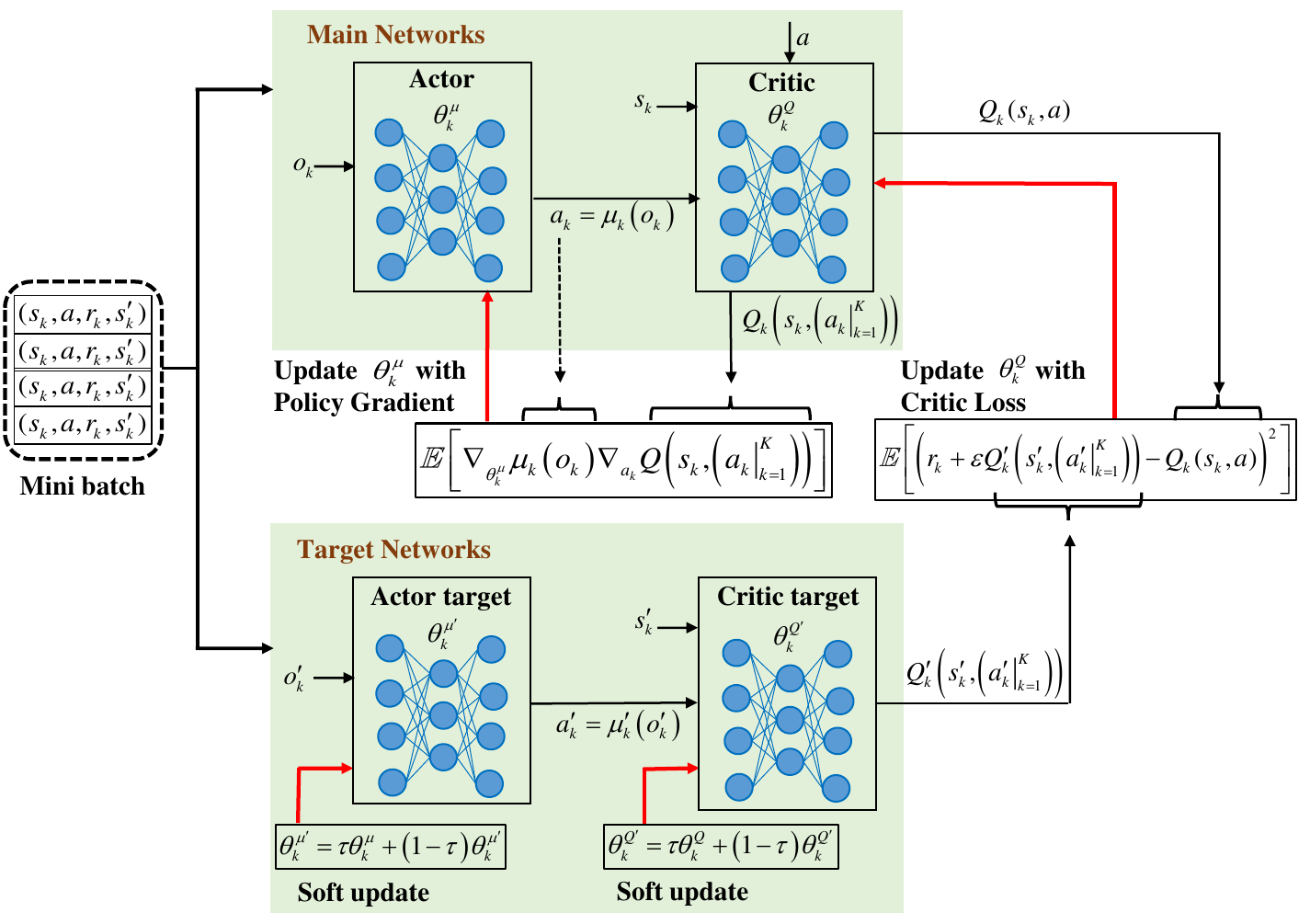}      
\caption{Illustration of the training procedure in MADDPG algorithm for the $k$-th agent}
\label{fig_sim}
\end{figure*}

Similar to its single agent counterpart, MADDPG is an actor-critic policy gradient algorithm. However, in the course of the offline training phase, the multi-agents are trained at a central unit which presents an opportunity for sharing extra information so as to ease the training process. As depicted in Fig. 3, in addition to the local observations of the environment, the agents are provided with additional information (shown as dotted lines), such as observations and actions of the other agents. Specifically, agent $k$  has now access to the joint action $a\left( t \right)=\left( {{a}_{1}}\left( t \right),...,{{a}_{K}}\left( t \right) \right)$, and full observation of the environment state ${{s}_{k}}\left( t \right)=\left( {{o}_{k}}\left( t \right),{{o}_{-k}}\left( t \right) \right)$, where ${{o}_{-k}}\left( t \right)$ is the local observations of other agents at time step $t$. The additional information, in particular the actions taken by other agents, can enable the multi-agents to overcome the challenges posed by the non-stationarity of the environment, thereby successfully capturing the dynamics of the mobile edge network. The extra information endowed to the agents during the centralized training phase, however, is discarded during the execution phase, meaning that the agents fully rely on their local observations to make JCCRA decisions in a fully distributed manner.

As shown in Fig. 4, the MADDPG agent $k$ employs two main deep neural networks: actor network with parameters $\theta _{k}^{\mu }$ to approximate a joint resource allocation policy ${{\mu }_{k}}\left( {{o}_{k}}|\theta _{k}^{\mu } \right)$ and a critic network, parametrized by $\theta _{k}^{Q}$, to approximate a state-value function ${{Q}_{k}}\left( {{s}_{k}},a|\theta _{k}^{Q} \right)$, along with their respective time-delayed copies, $\theta _{k}^{{{\mu }'}}$ and $\theta _k^{Q'}$ which serve as targets. At time step $t$, the actor deterministically maps local observation ${{o}_{k}}\left( t \right)$ to a specific continuous action ${{\mu }_{k}}\left( {{o}_{k}}\left( t \right)|\theta _{k}^{\mu } \right)$ and then, a random noise process ${{{\cal N}}_k}$ is added to generate an exploratory policy such that ${{a}_{k}}\left( t \right)={{\mu }_{k}}\left( {{o}_{k}}\left( t \right)|\theta _{k}^{\mu } \right)+ { {{\cal N}}_k}\left( t \right)$. The environment, which is shared among the agents, collects the joint action $a\left( t \right) = \left\{ {{a_k}\left( t \right),\forall k \in  {{\cal K}}} \right\}$ and returns the immediate reward ${{r}_{k}}\left( t \right)$ and the next observation ${{o}_{k}}\left( t+1 \right)$ to the respective agents. To make use of the experience in later decision-making steps, i.e., to improve sample efficiency, and stability of the training,  the agent’s transition along with the extra information ${{e}_{k}}\left( t \right)=\left( {{s}_{k}}\left( t \right),a\left( t \right),{{r}_{k}}\left( t \right),{{s}_{k}}\left( t+1 \right) \right)$ is saved in the replay buffer ${\cal D}_k$ of the agent $k$. 

To train the main networks, mini batch of $B$ samples $\left. {\left( {s_k^i,a_{}^i,r_k^i,s_k^{i + 1}} \right)} \right|_{i = 1}^B$, is randomly drawn from the replay buffer ${\cal D}_k$, denoting a sample index by $i$. The critic network is updated to minimize the following loss function:
\begin{equation}
{{\cal L}_k}\left( {\theta _k^Q} \right) = \frac{1}{B}\,\,\sum\nolimits_i {\,{{\left( {y_k^i - {Q_k}\left( {s_k^i,{a^i}|\theta _k^Q} \right)} \right)}^2}} \,\,,
\end{equation}
where $y_{k}^{i}$ is the target value expressed as $y_{k}^{i}={{\left. r_{k}^{i}+\varepsilon \,\,{{{{Q}'}}_{k}}\left( s_{k}^{i+1},{{a}_{i+1}}|\theta _{k}^{Q'} \right) \right|}_{{{a}_{^{i+1}}}=\left\{ {{{{\mu }'}}_{k}}\left( o_{k}^{i+1} \right),\forall k\in \cal{K} \right\}}}$, with a discount factor $\varepsilon \,$. More specifically, the parameters of the critic network, $\theta _k^Q$, are adjusted by following the gradient of (19) such that $\theta _k^Q \leftarrow \theta _k^Q - {\beta _Q}\,\,{\nabla _{\theta _k^Q}}{{\cal L}_k}\left( {\theta _k^Q} \right)$, with a learning rate of ${\beta _Q}$ for the critic network. On the other hand, the policy network  updates its parameters to maximize the expected long-term discounted reward $J\left( {{\mu }_{k}}|\theta _{k}^{\mu } \right)$ according to $\theta _{k}^{\mu }\leftarrow \theta _{k}^{\mu }-{{\beta }_{\mu }}\,\,{{\nabla }_{\theta _{k}^{\mu }}}J\left( {{\mu }_{k}}|\theta _{k}^{\mu } \right)$, where ${{\nabla }_{\theta _{k}^{Q}}}J\left( {{\mu }_{k}}|\theta _{k}^{\mu } \right)$ is the multi-agent deterministic policy gradient expressed as
\begin{multline}
	{\nabla _{\theta _k^\mu }}J\left( {{\mu _k}|\theta _k^\mu } \right) \approx \frac{1}{B}\,\Biggl[  \sum\nolimits_i {{\nabla _{{a_k}}}} {Q_k}\left( {s_k^i,{a_i}|\theta _k^Q} \right)
	\times  \\\nabla_{\theta _k^\mu }\,{\mu _k}\left( {o_k^i|\theta _k^\mu } \right) \Biggl]\Biggl|_{{a_i} = \left\{ {{\mu _k}\left( {o_k^i} \right),\,\forall k \in {{\cal K}}} \right\}} \, ,
\end{multline}
and ${{\beta }_{\mu }}$ is the learning rate of the actor network. The target parameters in both actor and critic networks are updated with soft updates as follows:
\begin{align}
& \theta _{k}^{{{\mu }'}}\leftarrow \tau \theta _{k}^{\mu }+\left( 1-\tau  \right)\theta _{k}^{{{\mu }'}} \nonumber \\ 
& \theta _{k}^{{{Q}'}}\leftarrow \tau \theta _{k}^{Q}+\left( 1-\tau  \right)\theta _{k}^{{{Q}'}},
\end{align}                    
where $\tau $ is a constant close to zero. 

\textcolor{black}{It is important to note that the agents are trained offline using the MADDPG algorithm, as summarized in Algorithm 1. The individual pre-trained actor models are employed for inference during the execution phase. In dynamic environments, the tracking capability of DRL algorithms empowers the agents to adapt to gradual shifts over time, such as minor variations in the user-configuration and task profiles by adjusting their policies based on their observations and perceived rewards. However, for more significant deviations from the training environment, such as sudden large influxes of users or increased shifts in task profiles, the pre-trained actor models may serve as a valuable starting point for performing on-the-fly fine-tuning. This involves training new models at the CPU using newly collected data from the changing environment to mitigate performance degradation caused by environmental drift, eliminating the necessity for retraining from the beginning.
\subsection{Complexity Analysis}
In this subsection, we analyze the computational complexity of the proposed algorithm during training and inference (execution) phases. The computational complexity during training is primarily determined by a structure of the two primary networks, namely the actor and critic. }

\textcolor{black}{
More specifically, for a fully connected actor and critic networks with $L$ layers, denoting the number of neurons in the $l$-th layer of the corresponding networks by  $N_{l}^{a}$ and  $N_{l}^{c}$, respectively, the complexity at each training step is given as $O\left( {\sum\nolimits_l {N_l^a} N_{l + 1}^a + \sum\nolimits_l {N_l^c} N_{l + 1}^c} \right),\,\forall l \in \left[ {0,L - 1} \right]$, where $l = 0$ denotes the respective input layers. Accordingly, the total computational complexity for training $K$ agents can be expressed as
\begin{equation}
O\left( {K{N_s}B\left( {\sum\nolimits_l {N_l^a} N_{l + 1}^a + \sum\nolimits_l {N_l^c} N_{l + 1}^c} \right)} \right),
\end{equation}
where ${{N}_{s}}$ represents the total number of training steps, and $B$ is the mini-batch size. On the other hand, the inference complexity of the agents solely depends on the individual actor networks, which can be given as $O\left( {\sum\limits_{l = 0}^{L - 1} {N_l^aN_{l + 1}^a} } \right)$. }

\section{Performance Analysis}
\subsection{Simulation Setup}
We start by uniformly distributing $M=100$ APs and $K=10$ active users over an area of 1km${^2}$. The APs are connected to the CPU via ideal fronthaul links. Unless stated otherwise, 30\% of the entire APs are clustered to serve each user, i.e.,  ${N_k} = 0.3M$, $k=1,2,\cdots ,10$. The system bandwidth is set to 5 MHz, which is shared among all users without channelization. A channel coefficient of the small-scale fading between the $k$-th user and $m$-th AP is set as \textcolor{black}{${h_{mk}} \sim {{\cal C}{\cal N}}\left( {0,1} \right)$}, which is independent across all APs and users. Furthermore, a channel gain of large-scale fading between the $k$-th user and $m$-th AP is given as
\begin{equation}
	{\beta _{mk}} = {10^{\frac{{P{L_{mk}}}}{{10}}}}{10^{\frac{{{\sigma _{sh}}{z_{mk}}}}{{10}}}},
\end{equation}
where ${\sigma _{sh}}$ represents the standard deviation of the shadow fading,  and ${z_{mk}}\!\sim \! {{\cal N}}\left( {0,1} \right)$. According to the three-slope model \textcolor{black}{[41]}, with fixed distance of ${{d}_{0}}$, and ${{d}_{1}}$, the path loss $P{{L}_{mk}}$ between the $k$-th user and $m$-th AP with a distance of ${d_{mk}}$ apart at the carrier frequency of $f$  is given as
\begin{equation}
	P{L_{mk}} =
	\begin{cases}
		- L - 35{\log _{10}}\left( {{d_{mk}}} \right) & \text{if 
			${d_{mk}} > d_1$}\\
		- L - 10{\log _{10}}\left( {d_{mk}^2\,d_1^{1.5}} \right) & \text{if ${d_0} < {d_{mk}} \le {d_1}$}\\
		- L - 10{\log _{10}}\left( {d_0^2\,d_1^{1.5}} \right) & \text{if ${d_{mk}} \le {d_0}$}
	\end{cases},      
\end{equation}
where 
\begin{multline}
	L = 46.3 + 33.9\,{\log _{10}}\left( f \right) - 13.82\,{\log _{10}}\left( {{h_{AP}}} \right) \\- \left( {1.1\,{{\log }_{10}}\left( f \right) - 0.7} \right){h_u} +  {1.56\,{{\log }_{10}}\left( f \right) - 0.8} \,,
\end{multline}
with ${h_{AP}}$ and ${{h}_{u}}$ to denote the antenna height (in meters) of AP and user, respectively. Further, there is no shadowing unless  ${d_{mk}} > {d_1}$. The values of  ${{d}_{0}}$, and ${{d}_{1}}$  are set to 10 m, and 50m, respectively, similar to \textcolor{black}{[9]}. The edge server at the CPU has computation capacity of ${f^{\rm{CPU}}} = 100\,{\rm{GHz}}$, while all users are equipped with the same computation capacity of $f_{k}^{\max }=1\,\text{GHz}$.  Furthermore, each bit requires 500 clock cycles to be processed, i.e., ${{N}_{cpb}}=500$. A time step $\Delta t$  is set to 1ms, and each user generates a task with a random size that is uniformly distributed in the range of  2.5 to 7.5 kbits, corresponding to a rate of 2.5 Mbps to 7.5 Mbps, at the beginning of every time step $t$. Without loss of generality, each user demands the task to be processed within the same interval, i.e., $t_k^d \le \Delta t = 1\,\rm{ms}$, for every time step. Meanwhile, we assume that a channel does not change over each time step, i.e., ${\tau _c} = 1\,\rm{ms}$. The simulation parameters are summarized in Table II. 
\begin{table}[t]
	\caption{Simulation parameters}
	\vspace{-0.8mm}
	\begin{tabular}{p{0.10\linewidth}|p{0.55\linewidth}|p{0.19\linewidth}}
		\hline
		\bfseries \vspace*{0.001mm}Notation  & \bfseries \vspace*{0.001mm}Parameter &  \bfseries \vspace*{0.001mm}Value\\ 
		\hline \hline
		\vspace*{0.0001mm}${f^{\rm{CPU}}}$ & \vspace*{0.0001mm}Computation capacity of the server in the CPU  & \vspace*{0.0001mm}$100 \,\,{\rm{GHz}}$\\ \hline
		\vspace*{0.0001mm}$f_k^{\max }$ &\vspace*{0.0001mm} Maximum local computation capacity & \vspace*{0.0001mm}$1 \,\,{\rm{GHz}}$\\ \hline
		\vspace*{0.00001mm}${N_{cpb}}$ & \vspace*{0.00001mm}Number of cycles to process one bit task  & \vspace*{0.00001mm}$500\,{\rm{cycles}}/{\rm{bit}}$ \\ \hline
		\vspace*{0.0001mm}$\varsigma $ & \vspace*{0.0001mm}Effective switched capacitance of the \,\,devices  & \vspace*{0.0001mm}${10^{ - 27}}$ \\ \hline
		\vspace*{0.0001mm}$t_k^d$ & \vspace*{0.0001mm}Application deadline for user $k$  & \vspace*{0.0001mm}$1\,\,{\rm{msec}}$ \\ \hline
		\vspace*{0.0001mm}$W$ & \vspace*{0.0001mm}System bandwidth  & \vspace*{0.0001mm}$5\,\,{\rm{MHz}}$ \\ \hline
		\vspace*{0.0001mm}$p_k^{\max }$ & \vspace*{0.0001mm}Maximum uplink transmit power  & \vspace*{0.0001mm}$0.1\,\,{\rm{W}}$ \\   \hline
		\vspace*{0.0001mm}$f$ & \vspace*{0.0001mm}Carrier frequency & \vspace*{0.0001mm}$1.9\,\,{\rm{GHz}}$\\   \hline
		\vspace*{0.0001mm}${h_{AP}}$ & \vspace*{0.0001mm}AP antenna height & \vspace*{0.0001mm}$15\,\,{\rm{m}}$\\  \hline
		\vspace*{0.0001mm}${h_{u}}$ & \vspace*{0.0001mm}User antenna height & \vspace*{0.0001mm}$1.65\,\,{\rm{m}}$\\  \hline
		\vspace*{0.0001mm}${\sigma _{sh}}$ & \vspace*{0.0001mm}Standard deviation of the shadow fading & \vspace*{0.0001mm}$10\,{\rm{dB}}$\\  \hline
		\vspace*{0.0001mm}${T_o}$ & \vspace*{0.0001mm}Noise temperature & \vspace*{0.0001mm}$290\,\,{\rm{K}}$\\  \hline
		\vspace*{0.0001mm}${\eta _f}$ & \vspace*{0.0001mm}Noise figure & \vspace*{0.0001mm}$9\,\,{\rm{dB}}$\\  \hline
		\vspace*{0.0001mm}${K_B}$ & \vspace*{0.0001mm}Boltzmann constant & \vspace*{0.0001mm}$1.381 \times {10^{ - 23}}\,\rm{J/K}$\\   
		\hline
	\end{tabular}
\end{table}
\begin{table}[t]
	\caption{Simulation hyperparameters for MADDPG training}
	\label{table_example}
	\centering
	\vspace{-0.3mm}
	\begin{tabular}{c|c}
		\hline
		\bfseries Hyperparameter &\bfseries Value  \\
		
		\hline\hline
		Discount factor $\varepsilon $ & 0.99 \\ 
		Soft update rate $\tau $ &0.005 \\
		Critic learning rate & 0.001 \\
		Actor learning rate & 0.0001 \\
		Mini batch size &128 \\
		Replay buffer size& 10000    \\
		\hline
	\end{tabular}
\end{table}

For the proposed MADDPG-based JCCRA, both actor and critic networks of each agent are implemented with fully connected neural networks, which consist of three hidden layers,  with 128, 64, and 64 nodes, respectively. All the hidden layers are activated by ReLu function, while the outputs of the actors are activated by sigmoid function. The target networks at each agent are copies of the respective actor and critic networks. The parameters of critic and actor networks are updated with adaptive moment (Adam) optimizer with learning rate 0.001 and 0.0001. Further, the discount factor and target networks update parameter are set to $\textcolor{black}{\varepsilon} = 0.99$ and  $\tau =0.005$, respectively, while the size of a mini batch is set to 128. Table III summarizes the hyperparameter values used in our simulations. The agents are trained with MADDPG algorithm for 1000 episodes, each consisting of 100 steps. In the next subsection, we present simulation results averaged from multiple experiments.  
\subsection{Performance Evaluation}
In this section, we present the performance of the proposed distributed MADDPG-based JCCRA. For performance comparison, we consider the following three benchmarks: 
\begin{itemize}
	\item  \textit{ \textbf {Centralized DDPG-based JCCRA scheme}}: This approach refers to DDPG-based centralized resource allocation scheme at the CPU. We adopt the same neural network structure and other hyperparameters as the MADDPG scheme to train the actor and critic in the single DDPG agent. However, the agent has full observation of the environment state, given in dimension of  $K\times 3$, during training and execution. The action has dimension of $K\times 2$, with each row corresponding to joint resource allocation for a particular user. Since global information of the entire network is processed centrally at the CPU to make JCCRA decisions for all users, we can potentially obtain the most efficient resource allocation. As a result, we intend to demonstrate a competitive performance of the distributed scheme against this baseline, which serves as a \emph{target} performance benchmark. However, as discussed in the previous sections, due to the additional time cost and associated overheads for two-way information exchange, this scheme may be infeasible to support time-sensitive applications. 
	\item  \textit{ \textbf {Offloading-first with an uplink fractional power control (FPC) scheme}}: This approach preferably offloads the computation to the edge server with the aim of aggressively exploiting the reliable access link provided by cell-free massive MIMO while saving the local processing energy consumption. The uplink transmit power for the $k$-th user is given by the standard fractional power control (FPC) \textcolor{black}{[42]} as follows:
	\begin{equation}
		{p_k} = \min \left( {p_k^{\max },{p_0}\lambda _k^{ - \nu }} \right),
	\end{equation} 
	where ${{p}_{0}}=-35\,\text{dBm}$, ${{\lambda }_{k}}=\sum\limits_{m=1}^{{{N}_{k}}}{{{\beta }_{mk}}}$, and $\nu =0.5$.
	\item  \textit{ \textbf {Local execution-first with an uplink fractional power control (FPC) scheme}}: The entire local processing capability is fully utilized, i.e., $f_{k}^{\rm{local}}=f_{k}^{\max }$, and the remaining task bits are offloaded to the edge server with uplink transmit power given according to \textcolor{black}{(26)}.
\end{itemize}

\textcolor{black}{We note that there is a lack of conclusive proof in the literature for the convergence of model-free actor-critic methods, or any other reinforcement learning algorithms using neural networks for function approximation. Consequently, we have empirically demonstrated the convergence of the proposed scheme in Fig. 5 by evaluating the total reward achieved over the training episodes, while comparing it against other benchmark schemes.} In contrast to the two heuristic approaches, it can clearly be observed that the learning-based schemes, i.e., centralized DDPG and the proposed MADDPG-based JCCRA, entertain higher reward values, which reflects lower total energy consumption while meeting the respective application deadlines of the users. The rewards per episode have smoothly increased in both learning schemes until they finally converge. A closer look at the convergence rate, however, reveals that the centralized scheme has converged faster than the distributed variant. This is attributed to the advantage of globally processing system wide information at the CPU to derive JCCRA policy for all users. Nevertheless, as the training episodes increase, the proposed algorithm has closed the gap and converged to the target benchmark, i.e., the centralized counterpart. \textcolor{black}{Moreover, as the agents are exposed to a larger number of scenarios in the stochastic MEC environment, they are able to learn and adapt with respect to the dynamics in the mobile edge network, leading to an efficient joint resource allocation policy with steady long-term performance.} It is also evident from Fig. 5 that conservative use of local computation resources leads to high energy consumption, while aggressively offloading results in ineffective utilization of the limited communication and computing resources.   
\begin{figure}[!t]
\captionsetup{singlelinecheck = false, justification=justified}
	\centering
	\includegraphics[width=3.5in,height=3.0in]{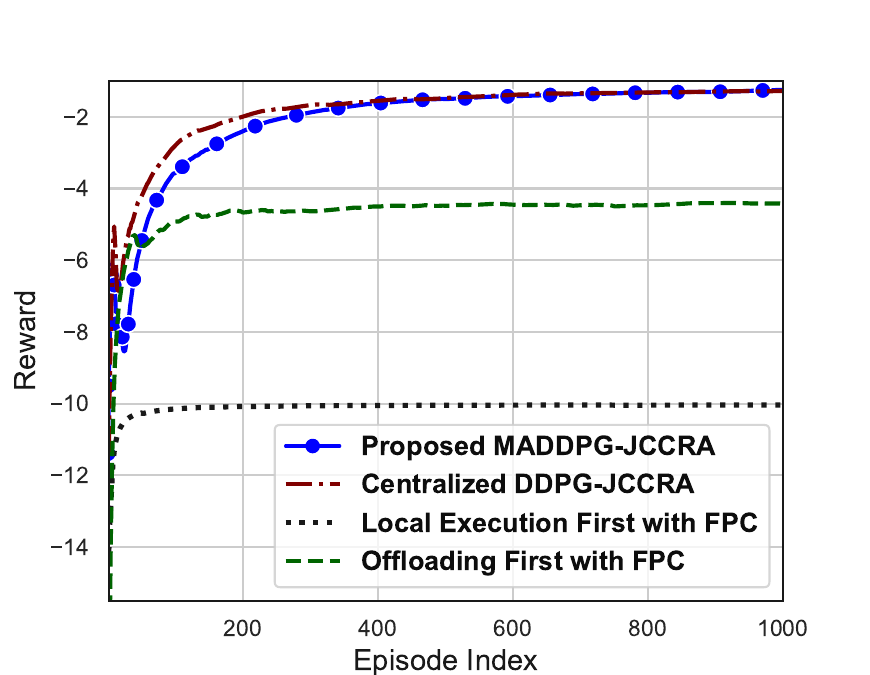}
	\caption{Total reward with training process}
	\label{fig_sim}
\end{figure}
\begin{figure}[t]
\captionsetup{singlelinecheck = false, justification=justified}
	\centering
	\includegraphics[width=3.5in,height=3.0in]{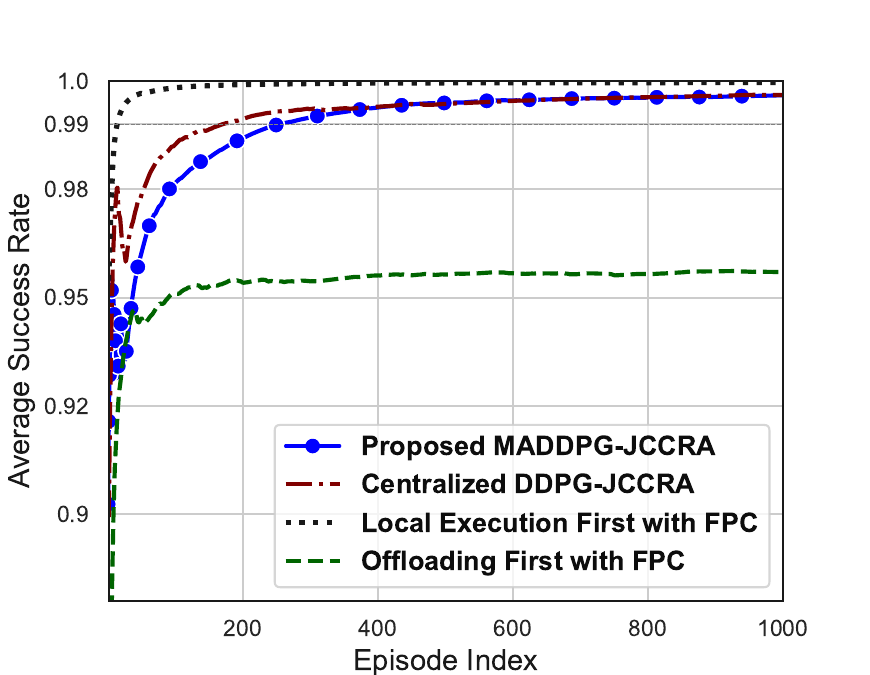}
	\caption{Average success rate with training process }
\end{figure}
\begin{figure}[t]
	\vspace*{-15pt}
	\centering
	\includegraphics[width=3.5in,height=3.0in]{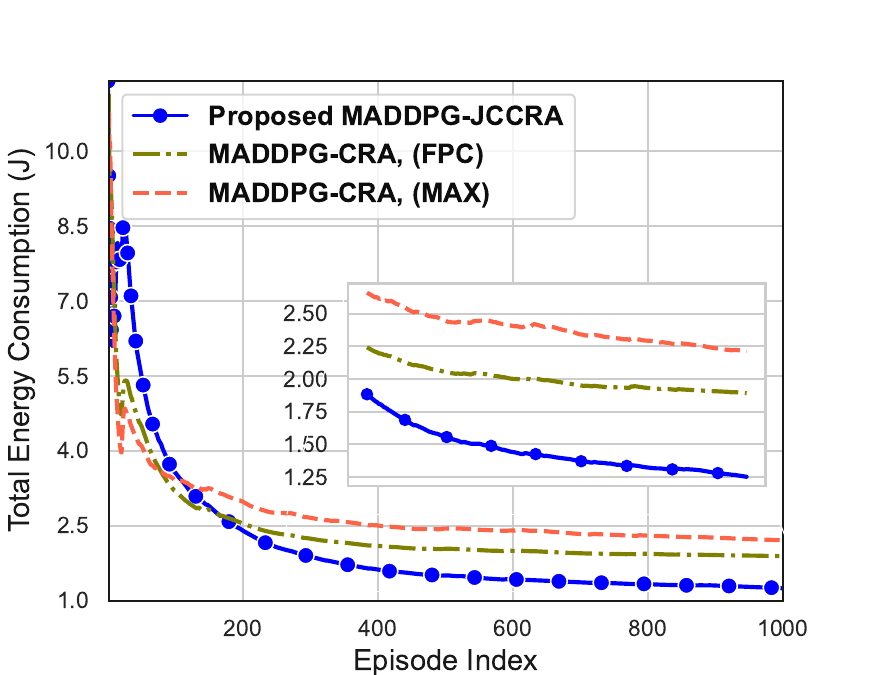}
	\caption{Comparison of the proposed JCCRA approach against variants with computing resource allocation optimization only}
\end{figure}

In Fig. 6, we compare the average success rate of the users in meeting the delay constraint of the time-sensitive applications. For the proposed MADDPG-based and centralized single-agent DDPG-based JCCRA schemes, the success rate under the simulation setup has gradually increased with episode indices to eventually reach more than 99\%, showing only a marginal gap from \emph{local execution-first} approach. The 100\% success rate in the latter scheme is, however, at the cost of unreasonably high energy consumption, which is 5 times more compared to the learning-based schemes, as observed in Fig 5. Moreover, the fact that the learning-based schemes outperform the conventional \emph{offloading-first} approach implies that even if cell-free massive MIMO can provide reliable access links to the users, aggressive computation offloading can result in performance degradation due to the ineffective use of the finite communication and computing resources, thereby subjecting users to service outage due to the failure to attain the tight delay constraints. It is, therefore, evident that an adaptive joint resource optimization is very critical to fully unleash the potential of the available communication and computing resources in the mobile edge network.
\begin{figure}[!t]
    \vspace*{-15pt}
	\centering
	\includegraphics[width=3.5in,height=3.0in]{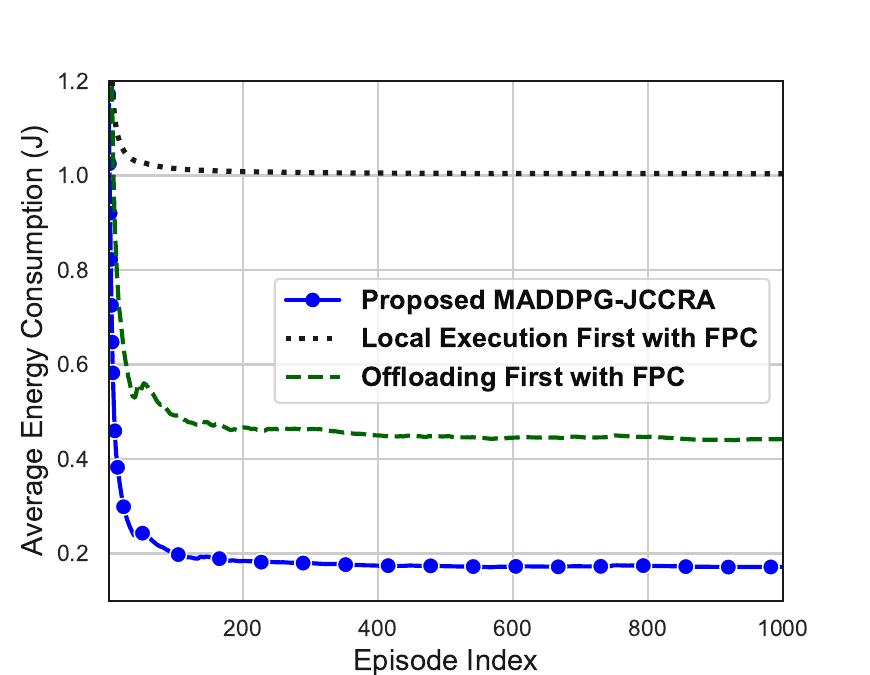}
	\caption{Per user average energy consumption at the testing (execution) stage}
\end{figure}

To further investigate a gain obtained by the joint resource allocation, we then compare the proposed JCCRA scheme with two variants that adopt MADDPG algorithm for optimizing computing resource allocation (CRA) only, i.e.  without taking power control into account. Additionally, we consider the heuristic fractional power control mechanism given in (25) for the first variant, referred to as MADDPG-CRA (FPC), while in the other variant, users offload computation with the maximum transmit power of ${p_k} = p_k^{\max },\forall k \in {{\cal K}}$, referred to as MADDPG-CRA (MAX). From Fig. 7, the performance degradation is readily observed in the variant schemes, which is reflected by 33\% and 42\% rise in total energy consumption of the user devices in  MADDPG-CRA (FPC), and MADDPG-CRA (MAX), respectively. This can be attributed to two factors. The first one is due to the critical role of power control in handling co-channel interference among different clusters during computation offloading. Thus, the adaptive MADDPG-based power control gives the proposed approach an edge over the variant schemes. The second factor is that computing and communication resource management in multi-user mobile edge network are inherently inter-dependent. Consequently, the proposed joint communication and computing resource allocation (JCCRA) optimization approach in the proposed scheme yields additional gain as compared with the separate optimization of computing resources in the other two variants.    
 
\begin{figure}[t]
\captionsetup{singlelinecheck = false, justification=justified}
  	\vspace*{-15pt}
	\centering
	\includegraphics[width=3.5in,height=3.0in]{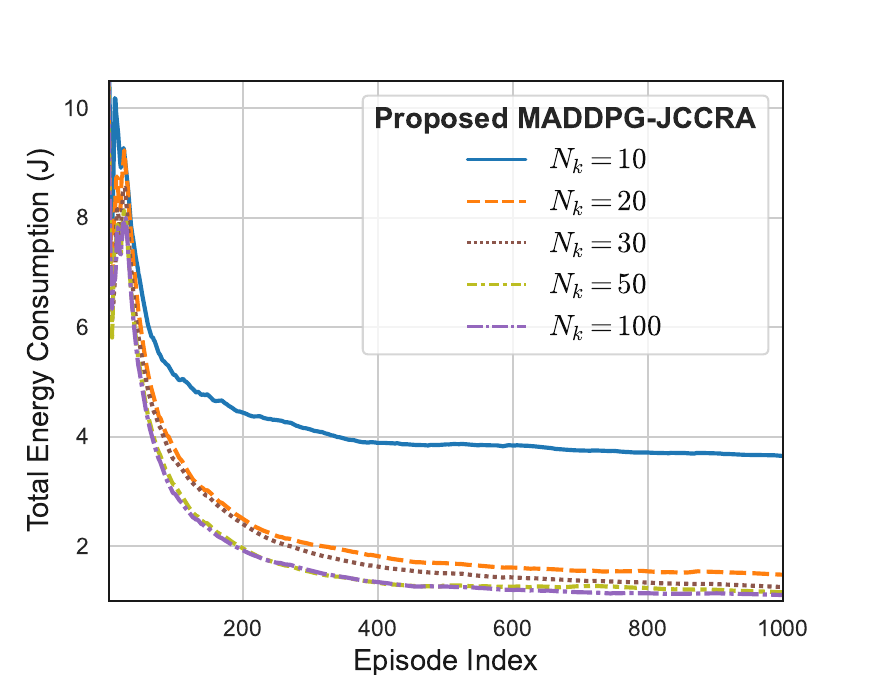}
	\caption{Effect of cluster size }
\end{figure}

After demonstrating the learning convergence and performance matching to the target centralized DDPG-based JCCRA scheme during the offline training stage, we then test the performance of the proposed JCCRA scheme in the considered MEC environment. Fig. 8 presents the per-user average energy consumption of the schemes during the execution stage. As it can be seen, the proposed approach achieves the lowest energy consumption, which is roughly 2.5 and 5 times lower than the \emph{offloading-first with FPC} and \emph{local execution-first with FPC} baseline schemes, respectively. We can deduce that the offline centralized training has successfully addressed the learning challenges of the multi-agent setting, enabling the agents effectively capture the complicated dynamics of the mobile edge network and perform well during the testing stage. Note that the execution phase is \emph{fully distributed} in the sense that the agents perform inference based on their local observations only, alleviating the burden associated with signaling overhead and additional delay in the centralized scheme. Specifically, in accordance with the learned policy, each MADDPG agent relies on its own actor network to map their local observation of the environment into efficient joint resource allocation in real time. 

Next, we investigate the effect of cluster size on the performance of the proposed framework. As discussed in Section III.A, we form a user-centric cluster of APs ${{{\cal C}}_k}$ to serve a given user by including ${{\rm{N}}_k}$ APs with the largest ${{\beta }_{mk}}$ (large-scale channel fading). In Fig. 9, we evaluate the average energy consumption by the different sizes of cluster, i.e., setting ${N_k} = 0.1M,0.2M,0.3M,0.5M,M$ for $M = 100$. It can be observed that the case of ${N_k} = 0.1M,\,\forall k \in {\cal K}$ has the highest energy consumption since the cell-free access rate is highly constrained to support computation offloading. The agents, therefore, have to either predominantly depend on local computation or increase the transmission power to meet the requirements of the multimedia applications, thus incurring high energy cost. \textcolor{black}{On the other hand, as ${{\rm{N}}_k}$ increases to $0.2M$ and $0.3M$, the performance improves notably, leading to a marginal gap compared to the canonical case where all APs serve every user, i.e., ${N_k} = 100$. In fact, the case of ${N_k} = 50$ has negligible performance gap in contrast to ${N_k} = 100$.} This indicates that with a sufficient number of APs in a cluster, a user-centric cell-free massive MIMO can effectively establish a reliable link for computation offloading with negligible performance loss. It should be noted that the user-centric approach requires a limited computational complexity for pilot detection and data processing, in contrast to the canonical counterpart. In practice, therefore, a legitimate cluster size must be set to deal with complexity and deployment cost while solving the JCCRA problem. 
\begin{figure}[!t]
	\centering
	\begin{subfigure}[b]{\linewidth}
		\centering
		\includegraphics[width=3.0in,height=2.6in]{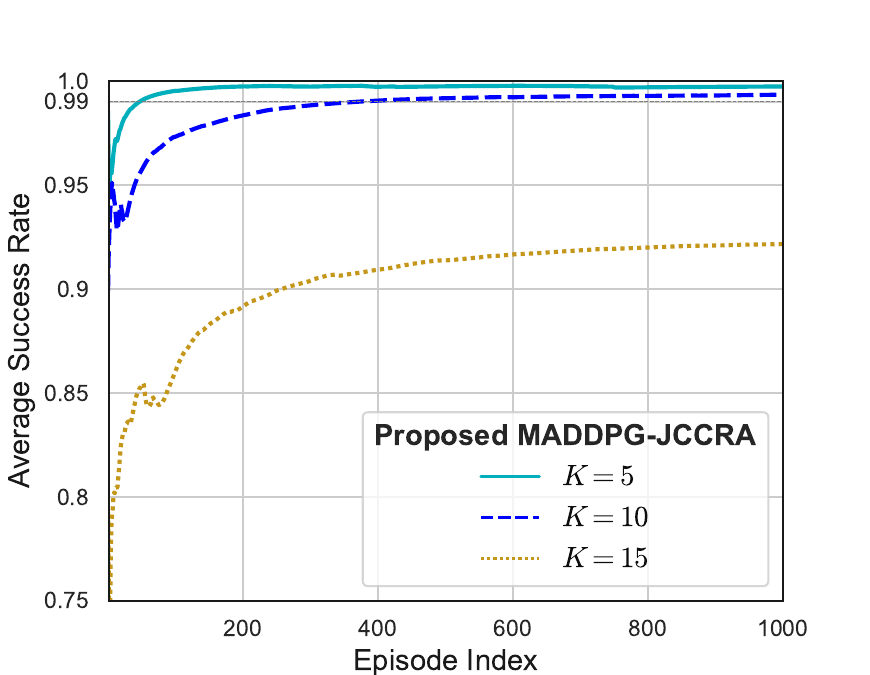}
		\label{fig_sim}
		\captionsetup{font=small,justification=centering}
		\caption{} 
	\end{subfigure}
	\begin{subfigure}[b]{\linewidth}
		\centering
		\includegraphics[width=3.0in,height=2.6in]{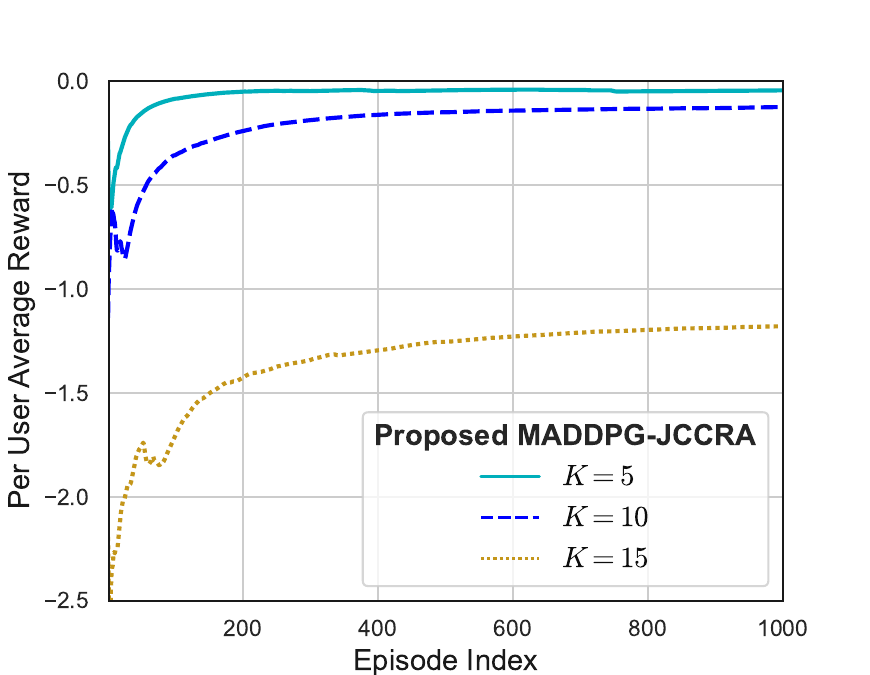}
		
		\label{fig_sim}
		\captionsetup{font=small,justification=centering}
		\caption{} 
	\end{subfigure}
	\captionsetup{font=small,justification=centering}
	\caption{Effect of the number of users in the mobile edge network. (a) Average success rate, (b) Per user average reward}
	\captionsetup{font=small,justification=justified}
 \vspace*{-10pt}
\end{figure}

The effect of the number of users on the performance of the proposed JCCRA scheme is investigated in Fig. 10, fixing $M = 100$ and ${N_k} = 0.3M$. One can observe that the proposed algorithm has converged in all cases. Since there are abundant communication and computing resources for a small number of users in the system, e.g., $K = 5$, the agents entertain almost 100\% success rate in accomplishing the intensive tasks within the deadline, while collecting the highest average reward, implying the lowest energy consumption compared to the other two cases. However, as the number of users increases, the competition among the users for limited communication and computing resources becomes more severe. In other words, the per-user resource reduces with an increasing number of users, e.g., the link rate decreases as $K$ increases. In particular, the case $K = 15$ represents a very constrained scenario that suffers from a significant outage. Herein, the available resources are not sufficient enough or the performance requirements are too strict to support all the devices within the stringent delay constraints, subjecting some users to service outage. This implies for critically constrained scenarios, either relaxing the performance requirements, for instance with soft delay constraints, or injection of more resources are required to support more users without outage. 

\begin{figure}[!t]
\captionsetup{singlelinecheck = false, justification=justified}
	\centering
	\begin{subfigure}[b]{\linewidth}
		\centering
		\includegraphics[width=3.0in,height=2.6in]{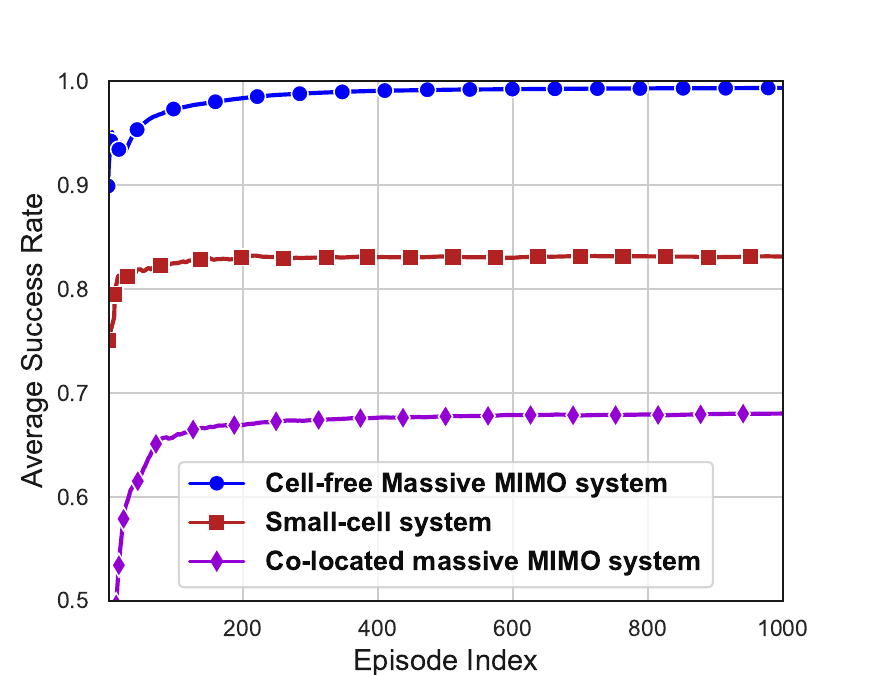}
		\label{fig_sim}
		\captionsetup{font=small,justification=centering}
		\caption{} 
	\end{subfigure}
	\begin{subfigure}[b]{\linewidth}
		\centering
		\includegraphics[width=3.0in,height=2.6in]{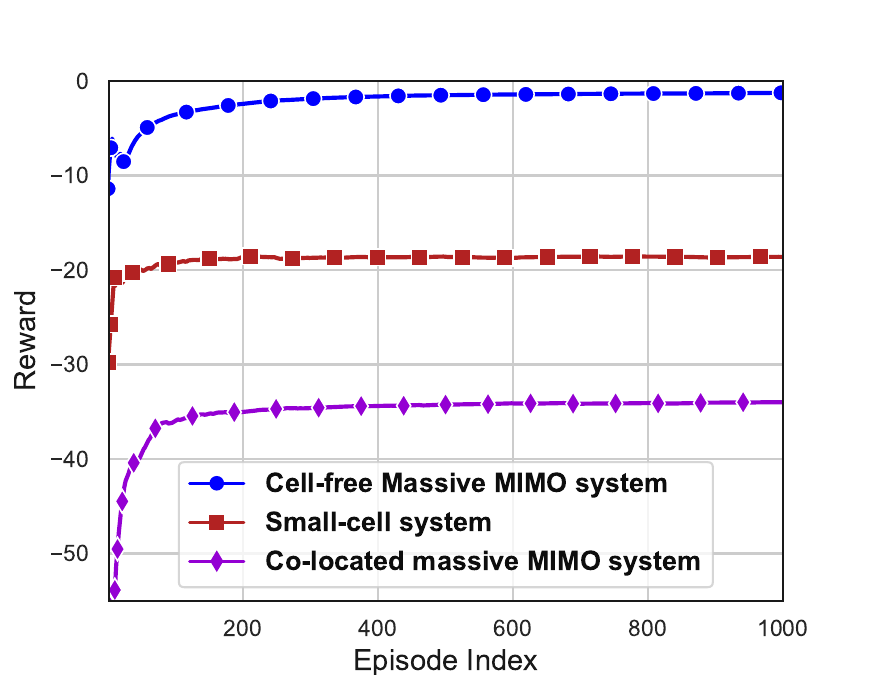}
		\label{fig_sim}
		\captionsetup{font=small,justification=centering}
		\caption{} 
	\end{subfigure}
	\captionsetup{font=small,justification=centering}
	\caption{Performance comparison of the proposed algorithm: Cell-free vs Cellular MEC systems. (a) Average success rate, (b) Total reward}
	\captionsetup{font=small,justification=justified, singlelinecheck=false}
	\vspace*{-10pt}
\end{figure}

Finally, we compare the performance of the proposed distributed approach in cell-free Massive MIMO with two cellular-based MEC systems, namely a small-cell system and a single-cell system with co-located massive MIMO. \textcolor{black}{For a fair comparison, a base station (BS) in the single-cell system is equipped with ${{N}_{k}}$ antennas, matching the number of APs in the cell-free system to simultaneously serve $K$ active users. Moreover, each user connects to a single AP with the largest large-scale fading coefficient, i.e., one with index  $m=\underset{m\in \left\{ 1,2,\cdot \cdot \cdot M \right\}}{\mathop{\arg \max }}\,{{\beta }_{mk}}$, in the small-cell system, following the methodology in [9]. Furthermore, to render edge computing services, both types of cellular MEC systems employ an edge server with a computation capacity of  ${{f}^{E}}={{f}^{\rm{CPU}}}$, where ${{f}^{\rm{CPU}}}$ corresponds to the edge server capacity in the cell-free MEC setup. Moreover, within the context of single-cell mobile edge computing (MEC), the edge server is located in close proximity to the base station (BS). Conversely, in the small-cell MEC system, to ensure a fair comparison with the cell-free MEC model, it is presumed that the edge server is connected to each access point (AP) through error-free links. In Fig. 11, we evaluate the performance of the proposed distributed approach in both cell-free and cellular MEC systems,  for $K=10$, ${{N}_{k}}=30$, and  ${{f}^{E}}={{f}^{\rm{CPU}}}=100\,\text{GHz}$. Furthermore, consistent computational and radio parameters are employed across all types of MEC systems in our simulations, as outlined in Table II. Demonstrated in} Fig.’s 11(a) and (b), the proposed algorithm performs significantly better \textcolor{black}{on a} cell-free system, in terms of average success rate and total reward accumulated, which also reflects the least total energy consumption according to (18), compared to the cellular MEC systems. The significant interference incurred from the co-located antennas under maximum ratio combining (MRC) in the single-cell co-located massive MIMO system, and the spatial co-channel interference from adjacent cells in the small-cell system has adversely affected computational task offloading by limiting the uplink rate. However, benefited from the distributed reliable links, the algorithm in cell-free MEC is shown to provide the most energy-efficient and consistently-low computational task offloading in contrast to the cellular MEC systems, shedding a light on the potential of our framework to handle the stringent requirements of advanced multimedia applications.  

\section{Conclusion and Future Works}
In this paper, motivated by its capability of realizing a reliable access link without cell edge, we presented a cell-free massive MIMO-enabled mobile edge network to address the stringent requirements of the next generation advanced services. We formulated a user-centric joint communication and computing resource allocation (JCCRA) problem to minimize the total energy consumption of the users while satisfying the corresponding delay constraints. We then proposed a distributed solution approach based on cooperative multi-agent reinforcement learning framework, wherein each user is implemented as a learning agent to make joint resource allocation relying only on local information during execution. The simulation results demonstrate that our distributed approach has outperformed the heuristic baselines, while matching the performance of the target centralized DDPG-based benchmark, without resorting to additional overhead and time cost. Furthermore, the distributed JCCRA in cell-free massive MIMO system is shown to provide a substantial performance improvement in terms of energy-efficiency and consistently-low latency task execution, compared to cellular MEC systems. Such a fully distributed and adaptive framework, enabled by cell-free massive MIMO system, can be a promising tool to realize edge intelligence for supporting the envisaged applications in dynamic mobile edge network. \textcolor{black}{We note that the current framework does not fully capture certain network dynamics present in practical MEC systems, including non-stationary scenarios characterized by significant variations in a user-configuration (e.g., distribution or mobility patterns) and task profiles (such as priority levels or task size distribution). We believe that such limitations can be addressed by incorporating techniques that can enhance generalization capability and fast adaptability, such as attention mechanism [43], meta learning [44], and transfer learning [45], which will be investigated in our future work. It would also be interesting to extend the current framework by incorporating mechanisms for dynamically determining a set of access points (APs) that can adapt to user mobility pattern. One possible approach is to dynamically configuring a cluster size of users in response to their mobility while taking into account the computing requirements [46].} Furthermore, for applications with a relatively relaxed delay tolerance, spanning multiple time steps, long-term performance optimization under queue dynamics can be another possible future direction.




\end{document}